\documentclass[12pt,aps,pre,preprint,showpacs,showkeys]{revtex4} 

\usepackage{graphicx}
\usepackage{subfigure}
\usepackage{amsmath}
\usepackage{amssymb}
\usepackage{bm}

\begin{document}

\title{XY model with competing higher-order interactions}
\author{Milan \v{Z}ukovi\v{c}}
 \email{milan.zukovic@upjs.sk}
\author{Georgii Kalagov\footnote{Current address: Joint Institute for Nuclear Research, Joliot-Curie, 6, Dubna 141980, Russia}}
 \affiliation{Institute of Physics, Faculty of Science, P. J. \v{S}af\'arik University, Park Angelinum 9, 041 54 Ko\v{s}ice, Slovakia}
\date{\today}

\begin{abstract}

We study effects of competing pairwise higher-order interactions (HOI) with alternating signs and exponentially decreasing intensity on critical behavior of the XY model. It is found that critical properties of such a generalized model can be very different from the standard XY model and can strongly depend on whether the number of HOI terms is odd or even. Inclusion of any odd number of HOI terms results in two consecutive phase transitions to distinct ferromagnetic quasi-long-range order phases. Even number of HOI terms leads to two phase transitions only if the decay of the HOI intensities is relatively slow. Then the high-temperature transition to the ferromagnetic phase is followed by another transition to a peculiar competition-induced canted ferromagnetic phase. In the limit of an infinite number of HOI terms only one phase transition is confirmed, and under the conditions of fierce competition between the even and odd terms the transition temperature can be suppressed practically to zero.

\end{abstract}

\pacs{05.10.Ln, 05.50.+q, 64.60.De, 75.10.Hk, 75.30.Kz}

\keywords{XY model, Higher-order interactions, Square lattice, Competition, Canted phase}



\maketitle

\section{Introduction}

The standard 2D XY model is well known to undergo a single topological Berezinskii-Kosterlitz-Thouless (BKT) phase transition, due to the vortex-antivortex pairs unbinding~\cite{bere71,kost73}. Below the BKT transition temperature the system displays a quasi-long-range-order (QLRO) characterized by a power-law decaying correlation function.


Nevertheless, its numerous modifications and generalizations, mostly obtained by including higher-order interaction (HOI) terms into the Hamiltonian, have produced a rather rich critical behavior with a number of different phases and a variety of the phase transitions~\cite{lee85,kors85,carp89,shi11,hubs13,qi13,pode11,cano14,cano16}. In particular, the well studied system with the Hamiltonian 
\begin{equation}
\label{Hamiltonian_J1-Jq}
{\mathcal H}=-J_1\sum_{\langle i,j \rangle}\cos(\phi_{i,j})-J_q\sum_{\langle i,j \rangle}\cos(q\phi_{i,j}),
\end{equation}
where $\langle i,j \rangle$ denotes the sum over nearest-neighbor spins and $J_1,J_2>0$ are the coupling constants, is known to show for $q=2$ separate dipole and quadrupole QLRO phases with the phase transition belonging to the Ising universality class~\cite{lee85,kors85,carp89,shi11,hubs13,qi13}. 

A more recent series of papers considered a nematiclike coupling of the order $q>2$ and found that the phase diagram topology can change for $q>3$~\cite{pode11,cano14,cano16}. The newly found phases were concluded to originate from the competition between the ferromagnetic $J_1>0$ and pseudonematic $J_q>0$ couplings and the identified phase transitions were demonstrated to belong to the 2D Potts, Ising, or BKT universality classes. These findings stirred theoretical interest as they pointed to a rather surprising lack of universality in systems showing the same $\phi \rightarrow \phi + 2\pi$ symmetry. However, the models that include HOI are also interesting from the experimental point of view. They have been shown to be applicable for modeling several physical (liquid crystals~\cite{lee85,geng09}, superfluid A phase of $^3{\rm He}$~\cite{kors85}, and high-temperature cuprate superconductors~\cite{hlub08}) as well as non-physical (DNA packing~\cite{grason08} and structural phases~\cite{cairns16,zuko16}) systems.

Some of the above experimental realizations involve HOI with negative signs. Unfortunately, such models that would include different HOI signs have received much less attention so far. They may include the magnetic interaction $J_1$, with collinear (ferromagnetic $J_1>0$ or antiferromagnetic $J_1<0$) ordering tendencies, and the generalized antinematic interactions $J_q<0$, which would favour noncollinear ordering. The coexistence of these antagonistic interactions induce competition between them, which may lead to the formation of novel phases~\cite{hayd10}. Indeed, the ferromagnetic-antinematic model, with $J_1>0$ and $J_2<0$, has been shown to feature at low temperatures another peculiar canted ferromagnetic (CF) QLRO phase, wedged between the ferromagnetic and antinematic phases~\cite{dian11,zuko19}. The new CF phase is characterised by ferromagnetic correlations which are significantly diminished by the presence of zero-energy domain walls due to the inherent degeneracy caused by the antinematic interactions. 

Besides the competition between the HOI terms, the model (\ref{Hamiltonian_J1-Jq}) can also involve some kind of frustration, which further adds to the model's complexity. As a result of its presence, the model on a square lattice with a frustration parameter shows an additional chiral phase transition, which occurs above (below) the BKT transition line if the magnetic and nematic couplings are of very different (comparable) strengths~\cite{qin09}. A similar phase diagram is also obtained for the geometrically frustrated model on a triangular lattice, however, with the separate chiral phase transition occurring above the BKT transition temperature regardless of the coupling strength ratio~\cite{park08}. Very recently it has been shown that in the model on the frustrated triangular lattice with $J_1<0,J_q<0$ and $q \geq 2$ the parameter $q$ can have a pronounced effect on the phase diagram topology~\cite{lach20,lach21}. In particular, for values of $q$ divisible by 3, the ground-state competition between the two interactions resulted in a new frustrated canted antiferromagnetic phase, appearing at low temperatures, and the increasing $q$ nondivisible by 3 led to the evolving phase diagram topology featuring two ($q = 2$), three ($q = 4, 5$), and four ($q \geq 7$) ordered phases.

In some of our previous studies we further generalized the model (\ref{Hamiltonian_J1-Jq}) by taking into account effects of more HOI terms with purely positive signs, including an infinite number~\cite{zuko17,zuko18a,zuko18b}. Among others, we found the change to the first-order transition in the models involving a sufficiently large number of HOI terms with an exponentially vanishing strength~\cite{zuko17} or the possibility of multiple phase transitions between different phases in the models including only a small (2-3) number of HOI terms for some setting of their strengths~\cite{zuko18a,zuko18b}.

In the present investigations we study the generalized XY model, which includes up to infinite number of HOI terms with alternating signs. The coexistence of such HOI terms can result in non-trivial ways of their mutual competition and collaboration in establishing some kind of preferred ordering. Besides the importance of HOI, demonstrated in the above studies, considering such a model is also motivated by some realistic systems in biology (DNA packing~\cite{grason08}) and chemistry (structural phases of cyanide polymers~\cite{cairns16,zuko16}). Its application can be facilitated by an appropriate mapping between magnetic interactions in the generalized XY model and (supra)molecular forces, which can have in the respective systems attractive or repulsive characters. It is often assumed that the essential features of the system behavior can be captured by considering only the simplified interaction potential with the first or first two Fourier components of the real interaction potential. In this study we show that higher-order terms can also play an important role in correct identification of the present phases.

\section{Model}

We consider a collection of classical two-component spins (unit vectors) localized on a square lattice, which interact with their nearest neighbors through the pairwise potential
\begin{equation}
\label{Potential}
H_{i,j}(p,\alpha)=\sum_{k=1}^{p}J_{k}\cos^{k}\phi_{i,j},
\end{equation}
where $\phi_{i,j}=\phi_{i}-\phi_{j}$ is an angle between the neighboring spins and the constants $J_k$ represent weights of the respective (higher-order) terms in the summation. We assume that their intensity decays exponentially but their signs alternate between positive and negative values for $k$ odd and even, respectively, i.e., $J_k=-(-\alpha)^{-k}$, where $\alpha>1$. Then, the Hamiltonian can be expressed in the form
\begin{equation}
\label{Hamiltonian}
{\mathcal H}(p,\alpha)=J(p,\alpha)\sum_{\langle i,j \rangle}H_{i,j}(p,\alpha),
\end{equation}
where $J(p,\alpha)$ is a constant adjusted to reset the potential within the interval $[-1,1]$ and $\langle i,j \rangle$ denotes the sum over nearest-neighbor spins.

In the limiting case of $p \to \infty$, i.e., with the infinite number of the higher-order terms (hereafter IHOI model) the Hamiltonian simplifies to 
\begin{equation}
\label{Hamiltonian_inf}
{\mathcal H}(\alpha)=J(\alpha)\sum_{\langle i,j \rangle}H_{i,j}(\alpha)=-J(\alpha)\sum_{\langle i,j \rangle}\frac{\cos\phi_{i,j}}{\alpha+\cos\phi_{i,j}},
\end{equation}
where $J(\alpha)=\alpha-1$ is the exchange interaction constant chosen to rescale the weights $J_{k}$ to sum up to unity.

In the more general case when the number of the higher-order terms $p$ stays finite (hereafter FHOI model), the Hamiltonian takes the form
\begin{equation}
\label{Hamiltonian_fin}
{\mathcal H}(p,\alpha)=J(p,\alpha)\sum_{\langle i,j \rangle}H_{i,j}(p,\alpha)=-J(p,\alpha)\sum_{\langle i,j \rangle}\frac{\cos\phi_{i,j}\Big[\Big(-\frac{\cos\phi_{i,j}}{\alpha}\Big)^{p}-1\Big]}{\alpha+\cos\phi_{i,j}},
\end{equation}
where $J(p,\alpha)=(\alpha-1)/(\alpha^{-p}-1)$. 

\begin{figure}[t!]
\centering
\subfigure{\includegraphics[scale=0.4,clip]{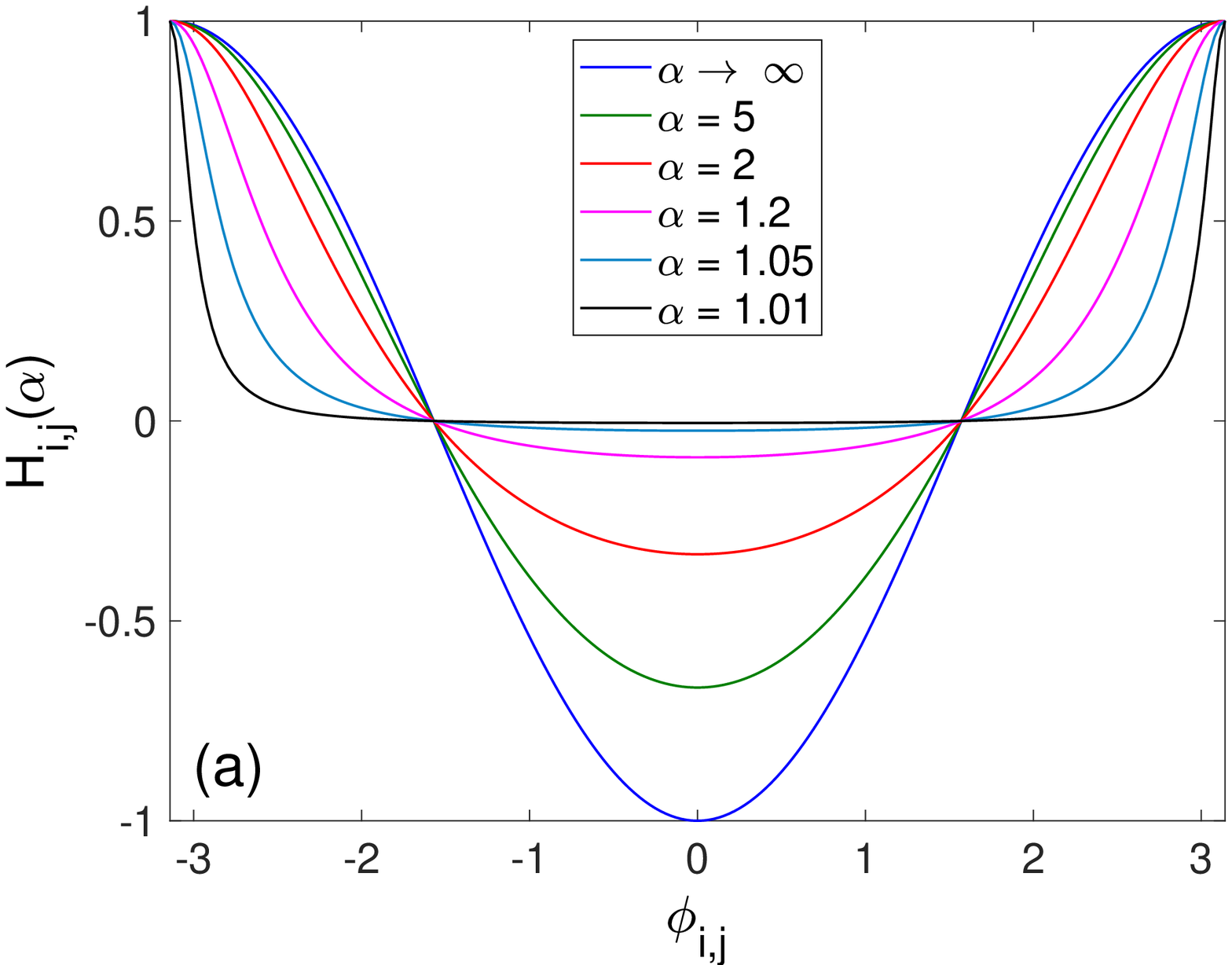}\label{fig:en_well_inf}} 
\subfigure{\includegraphics[scale=0.4,clip]{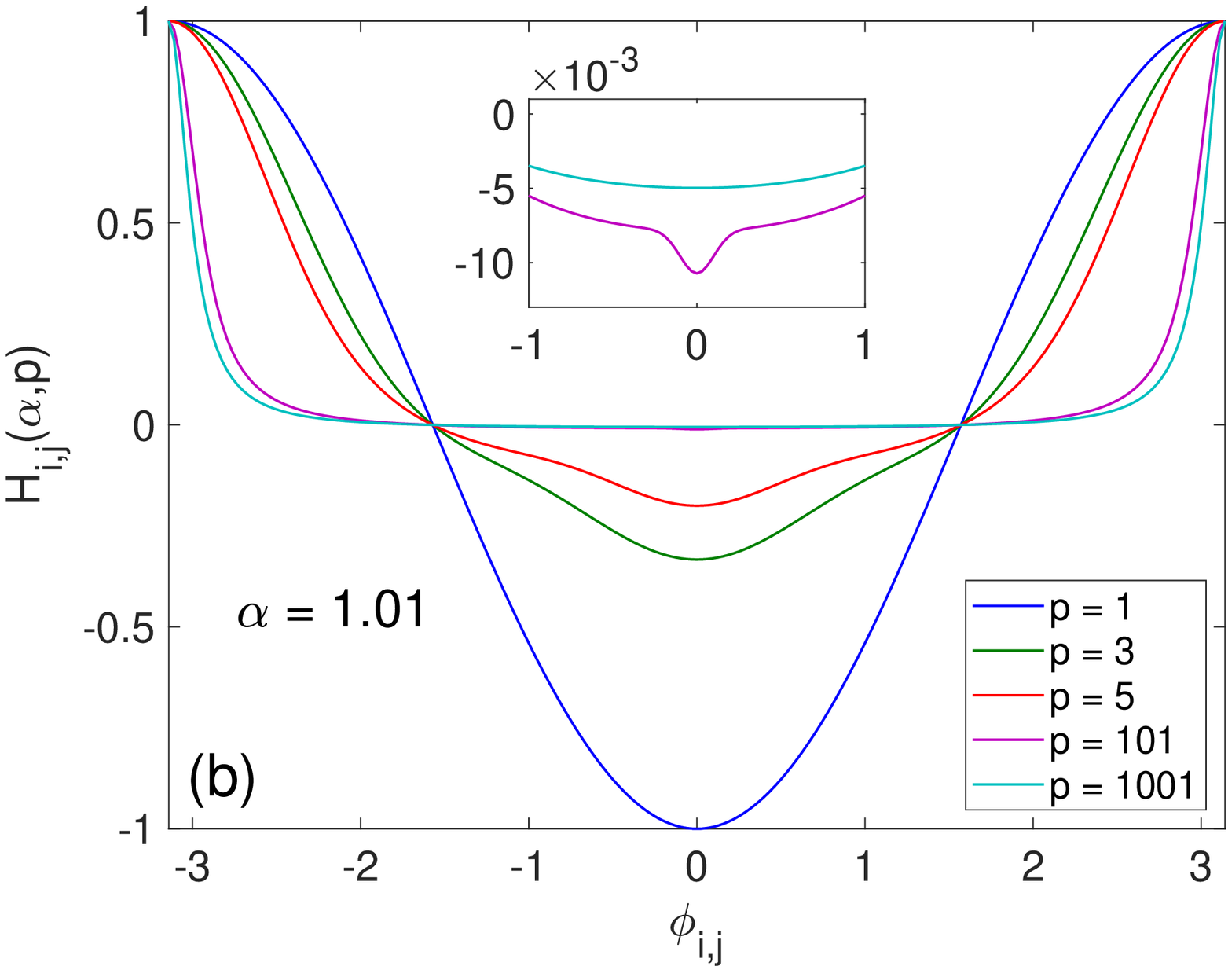}\label{fig:en_well_fin_p_odd_alp_001}}\\
\subfigure{\includegraphics[scale=0.4,clip]{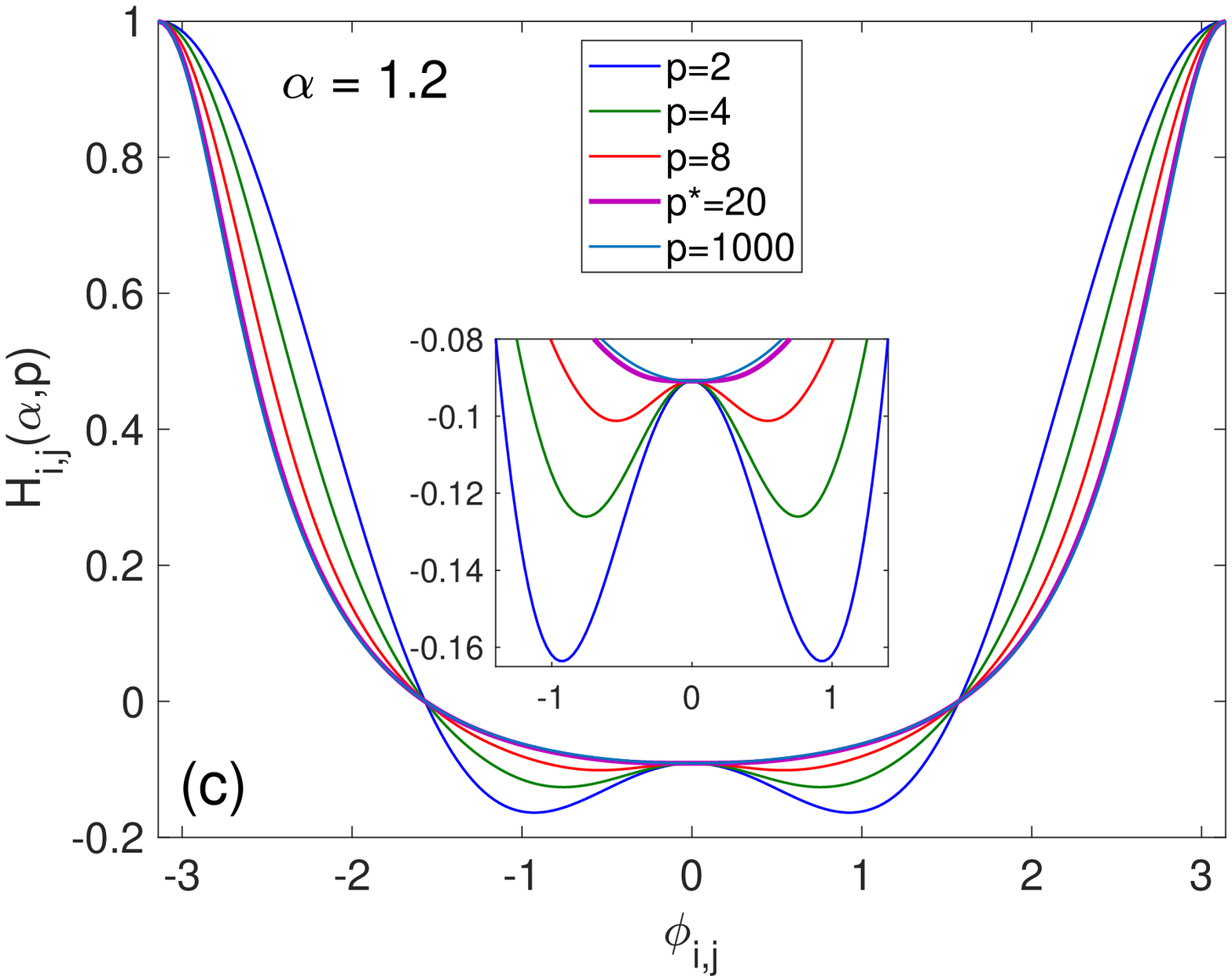}\label{fig:en_well_fin_p_even_alp_1_2}}
\subfigure{\includegraphics[scale=0.4,clip]{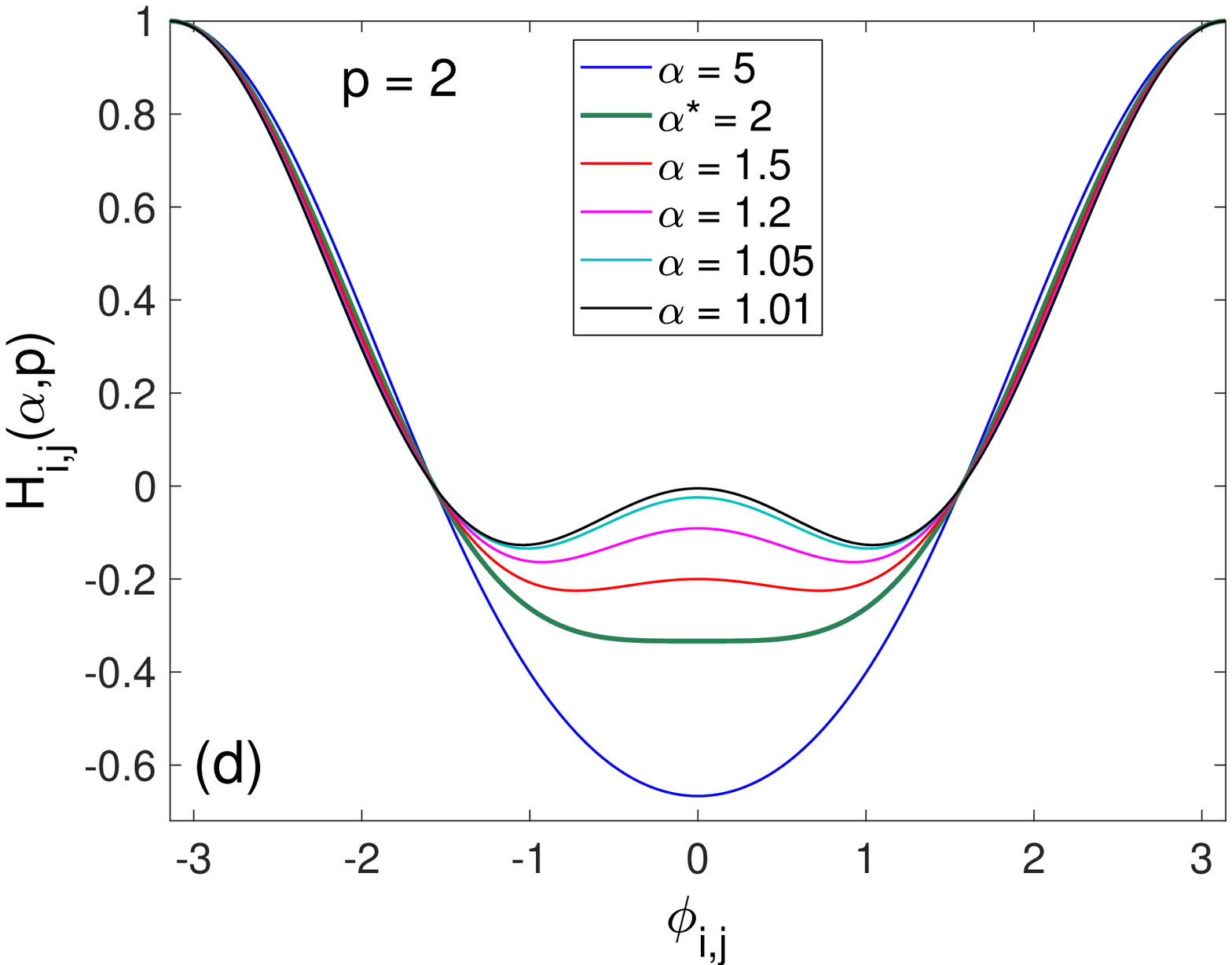}\label{fig:en_well_fin_p2}}
\caption{(Color online) Potential functions for the cases of (a) $p \to \infty$ for several values of $\alpha$, (b) the fixed $\alpha=1.01$ and various \emph{odd} values of $p$, (c) the fixed $\alpha=1.2$ and various \emph{even} values of $p$, and (d) the fixed $p=2$ and various values of $\alpha$. In (c) and (d) $p^*=20$ and $\alpha^*=2$ denote transition values from single- to double-well shape.}\label{fig:well}
\end{figure} 

Thus, there is one parameter, $\alpha$, and two parameters, $\alpha$ and $p$, which can be used to control the shapes of the potentials of  the IHOI and FHOI models, respectively. The shapes of the potential functions for various cases are presented in Fig.~\ref{fig:well}. The one in the IHOI model, shown in Fig.~\ref{fig:en_well_inf}, reduces to the conventional XY model when the interaction intensity decays extremely fast, i.e., for $\alpha \to \infty$. We note that for $\alpha \to \infty$ the value of $p$ is practically irrelevant and the shape very close to a cosine function would be also observed for any $p$ in the FHOI case. With the decrease in $\alpha$, the upper part of the potential well gets broader, the bottom rises towards positive values and is becoming flat with the width tending to $2\pi$ as $\alpha \to 1$. 

In the FHOI model, it is useful to distinguish between the cases with the odd and even number of the HOI terms. In the former case one can observe a similar effect of the increasing number of the HOI terms on the potential shape, for sufficiently small values of $\alpha$ (see Fig.~\ref{fig:en_well_fin_p_odd_alp_001}). In this case, the standard XY model is recovered for $p=1$ and with the increasing $p$ the potential well again gets broader in the range of positive values with the bottom becoming flat but a small bump at $\phi =0$ remains up to very large $p$ (see the inset). For even $p$ the potential function can show two distinct shapes, depending on the actual values of the parameters. In particular, for sufficiently small values of both $p$ and $\alpha$ the potential has a double-well shape, which is symmetrical around $\phi =0$. With the increasing either $p$ (see Fig.~\ref{fig:en_well_fin_p_even_alp_1_2}) or $\alpha$ (see Fig.~\ref{fig:en_well_fin_p2}) the double-well becomes shallower until it is converted to a single-well shape with the bottom at $\phi =0$ for some threshold value of $p^*$ or $\alpha^*$, respectively.



\section{Methods}

\subsection{Spin wave approximation}
We are interested in a large-scale behavior of the pair correlation function $ g_k(x_1-x_2) \equiv \left\langle \cos k [ \phi(x_1) -  \phi(x_2) ] \right \rangle $, where $k \in \mathbb{N}$ and brackets $\left\langle \dots  \right \rangle $ denote an average over all possible spin configurations with the Gibbs distribution functional  $\exp(-\beta {\mathcal H})$, where ${\mathcal H}$ is the Hamiltonian~(\ref{Hamiltonian}) and $\beta$ is the inverse temperature. Let $x$ be the coordinate vector of the $i$th spin and $a$ be the lattice vector. Then passing to the continuous limit within the spin-wave (SW) approximation, we arrive at the effective Hamiltonian 
 \begin{equation}
{\mathcal H_{\rm eff}} = \frac{J_{\rm eff}}{2} \int \mathrm{d}^2 x [\nabla \phi(x)]^2,
 \end{equation}     
where the effective coupling $J_{\rm eff} \equiv -\sum_{k=1}^{p} k/(- \alpha)^k$. Direct computation of the correlation function in the large-scale region $|x_1 - x_2| \gg a$ using the effective Hamiltonian
 \begin{equation}
g_k(x_1 - x_2) = \int \prod_{x} \mathrm{d} \phi(x) \exp\left( -\beta {\mathcal H_{\rm eff}} + i k [\phi(x_1)  - \phi_{x_2}]\right)
 \end{equation}
leads to the result 
 \begin{equation}
g_k(x_1 - x_2) = C_0 \exp\left( -\frac{ k^2}{2 \pi \beta J_{\rm eff}} \ln \frac{|x_1-x_2|}{a}\right)  \propto \left(\frac{a}{|x_1 - x_2|}  \right)^{\eta_{\rm eff}}.   
\label{g_k}  
 \end{equation}
 Here $C_0$ is an unessential constant and the critical exponent $\eta_{\rm eff} = k^2/(2 \pi \beta J_{\rm eff})$.  It is important to note that in order to perform the Gaussian integration correctly, the effective exchange interaction $J_{\rm eff}$ has to be a positive quantity. In the case at hand, one needs to separately consider the odd and even values of $p$. Indeed, the magnitude of $J_{\rm eff}$ reads 
\begin{equation}
J_{\rm eff} = \left\{
\begin{aligned}
 \frac{\alpha}{(1+\alpha)^2} - \frac{p \alpha + p +\alpha}{\alpha^p (1+\alpha)^2}, \quad & \text{for even}\, p, \\
 \frac{\alpha}{(1+\alpha)^2} + \frac{p \alpha + p +\alpha}{\alpha^p (1+\alpha)^2}, \quad & \text{for odd}\, p.
\end{aligned}
\right.
\label{J_eff} 
\end{equation}
As $p$ tends to infinity  the coupling $J_{\rm eff}$ approaches the limit $\alpha/(1+\alpha)^2$ being always positive. However, for even $p$ the amplitude $J_{\rm eff}$ as the function of $\alpha$ might vanish and become negative. The value $\alpha = \alpha^*$ at which $J_{\rm eff}$ turns to be zero meets the equation 
\begin{equation}
{\alpha^*}^{p+1} =  \alpha^*(p+1) + p.
\label{alp}
\end{equation}
This equation has a unique root $\alpha^* \in [1, 2]$, $\forall p \in 2 \mathbb{N}$, and the SW approximation remains valid for $\alpha > \alpha^*$. We note that the value of $\alpha^*$, for a given $p$, (or $p^*$ for a given $\alpha$) corresponds to the transition between the single- and double-well shapes of the potential, shown in Fig.~\ref{fig:well}.

\subsection{Monte Carlo simulation}
Standard Monte Carlo (MC) simulations following the Metropolis dynamics is applied to study spin systems on a square lattice of the size $L \times L$, with $L=24-120$ and the periodic boundary conditions. For thermal averaging we take $N_{MC}=2 \times 10^5$ MC sweeps after discarding another $N_0=0.2\times N_{MC}$ MC sweeps used for bringing the system to equilibrium. A simulation usually starts at a sufficiently high temperature $T$ (measured in units $J/k_B$, where $k_B$ is the Boltzmann constant), corresponding to the paramagnetic phase. Then the simulation proceeds to the lower temperature $T-\Delta T$, at which it is initialized using the last configuration obtained at $T$. Typically, several independent runs are carried out for $L=48$. 

We obtain temperature dependencies of the following quantities: the internal energy per spin $e=\langle {\mathcal H} \rangle/L^2$, 
the specific heat per spin $c$
\begin{equation}
c=\frac{\langle {\mathcal H}^{2} \rangle - \langle {\mathcal H} \rangle^{2}}{L^2T^{2}},
\label{c}
\end{equation}
the generalized magnetizations per spin
\begin{equation}
m_q=\langle M_q \rangle/L^2=\left\langle\Big|\sum_{j}\exp(iq\phi_j)\Big|\right\rangle/L^2,\ q=1,\hdots,p,
\label{m}
\end{equation}
and the corresponding susceptibilities
\begin{equation}
\label{chi}\chi_q = \frac{\langle M_q^{2} \rangle - \langle M_q \rangle^{2}}{L^2T}.
\end{equation} 
At the BKT transition from the paramagnetic phase, expected for the XY model, the standard magnetization $m_1$ vanishes as power law, characterized by the exponent $\eta=1/4$. The latter can be estimated within the BKT phase as a function of temperature by the finite-size scaling (FSS) analysis as follows
\begin{equation}
\label{m_FSS}
m_1(T,L) \propto L^{-\eta(T)/2},
\end{equation}
considering the lattice sizes $L=48-120$.

\section{Low-temperature behavior}

\begin{figure}[t!]
\centering
\subfigure{\includegraphics[scale=0.4,clip]{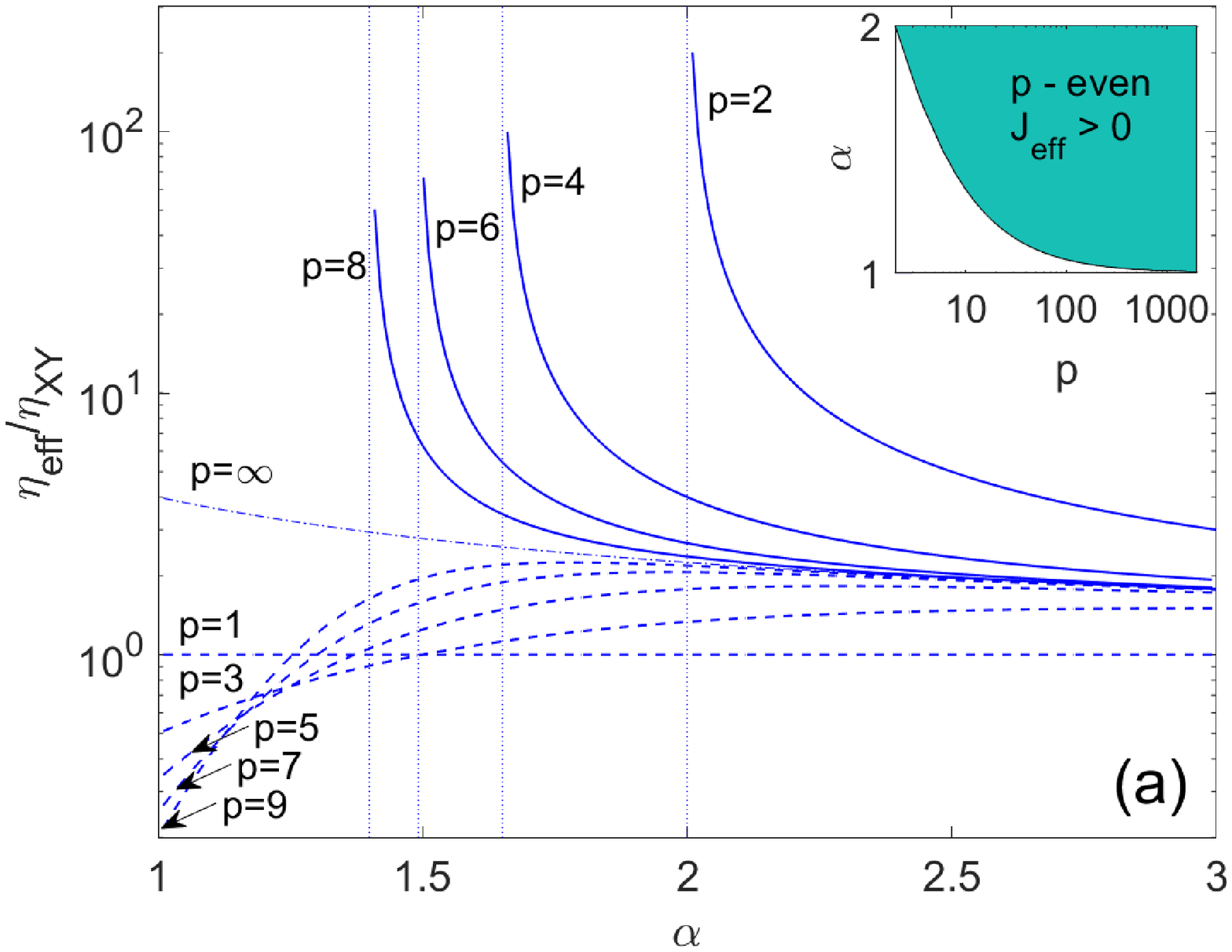}\label{fig:eta_eff-alp_new}}
\subfigure{\includegraphics[scale=0.4,clip]{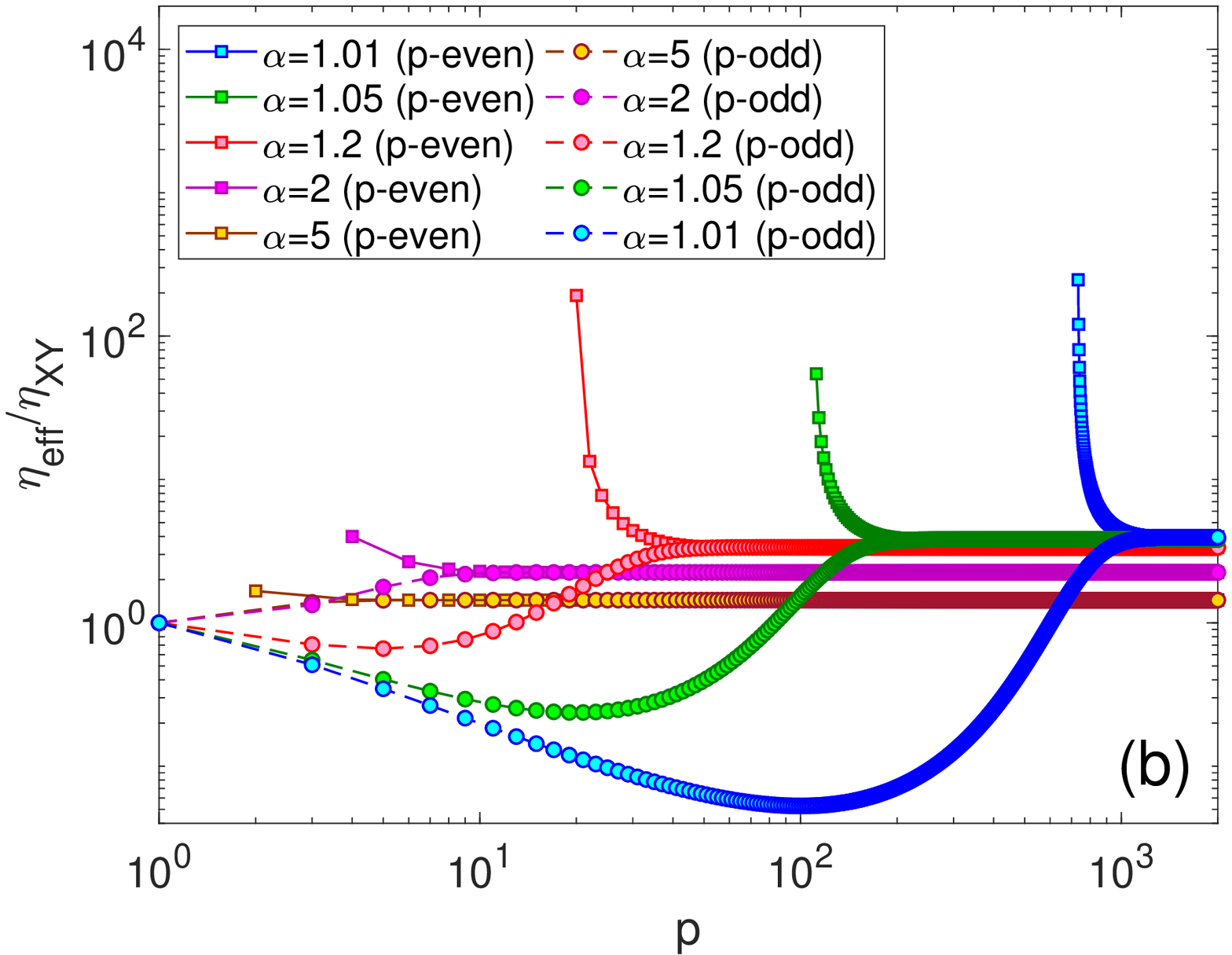}\label{fig:eta_eff-p_new}}
\caption{(Color online) (a) SW approximation of the reduced correlation function exponent $\eta_{\rm eff}/\eta_{\rm XY}$, shown as a function of (a) $\alpha$ for different $p$ and (b) $p$ for different $\alpha$. The solid and dashed curves represent the cases of even and odd values of $p$, respectively. In (a) the vertical broken lines denote the respective limiting values $\alpha^*(p)$ and the inset shows for even $p$ the area of $J_{\rm eff}>0$ in which $\eta_{\rm eff}$ is defined.}\label{fig:low-T}
\end{figure}

At low temperatures the QLRO phase in the SW approximation is characterized by the power-law decaying correlation function, given by Eq.~(\ref{g_k}). In contrast to the standard XY model with the critical exponent $\eta_{\rm XY}=T/( 2\pi J)$, in the present model the exponent $\eta_{\rm eff}$  also depends on the parameters $p$ and $\alpha$, involved in the interaction $J_{\rm eff}$, in a nontrivial way. In Fig.~\ref{fig:low-T} we present the reduced critical exponent $\eta_{\rm eff}/\eta_{\rm XY}=J_{\rm XY}/J_{\rm eff}$, where $J_{\rm XY}=J_{\rm eff}(p=1)=1/\alpha$, as a function of the parameters $\alpha$ (Fig.~\ref{fig:eta_eff-alp_new}) and $p$ (Fig.~\ref{fig:eta_eff-p_new}). As one can anticipate from the form of $J_{\rm eff}$ in Eq.~(\ref{J_eff}), the behaviors for odd and even $p$ will be different. 

For odd values of $p$ and small values of $\alpha \to 1$, the inclusion of the HOI terms causes a dramatic drop of the exponent down to $\eta_{\rm eff}/\eta_{\rm XY} \approx 0.053$ at $p \approx 100$, followed by a gradual increase and leveling off at the limiting value $\eta_{\rm eff}/\eta_{\rm XY} \approx (\alpha+1)^2/\alpha^2$ above $p \approx 10^3$. For larger $\alpha$ the minimum value of $\eta_{\rm eff}$ increases but it is reached at smaller values of $p$. Eventually, $\eta_{\rm eff}$ becomes a purely increasing function of $p$ above $\alpha \approx 1.5$.  

On the other hand, for even values of $p$ the critical exponent $\eta_{\rm eff}$ is only defined for $\alpha > \alpha^*$, where $\alpha^*$ is a solution of Eq.~(\ref{alp}). As demonstrated in Fig.~\ref{fig:eta_eff-alp_new}, for any $p$ the reduced exponent $\eta_{\rm eff}/\eta_{\rm XY}$ takes its maximum at $\alpha^*$ and then it gradually decreases to unity as $\alpha \to \infty$. A qualitatively similar behavior can be observed for a fixed value of $\alpha$ and $p$ being increased from $p^*$ to $\infty$ (Fig.~\ref{fig:eta_eff-p_new}). It is worth noting that for even number of terms $\eta_{\rm eff} \geq \eta_{\rm XY}$ in the whole $(p-\alpha)$ parameter plane. Finally, by comparing the reduced critical exponent for odd and even values of $p$ one can conclude that the former is always smaller than the latter, albeit, beyond some value of $\alpha$ or $p$ differences become negligible.

\section{Phase diagrams}
\subsection{FHOI model}
Below we study a finite-temperature behavior of the FHOI model with the goal to construct phase diagrams in the $(p-\alpha)$ parameter space. In particular, we calculate temperature dependencies of various quantities, which can provide locations of the transition temperatures (peaks in the response functions) and reveal the nature of the ordering (generalized magnetizations).
 
\begin{figure}[t!]
\centering
\subfigure{\includegraphics[scale=0.55,clip]{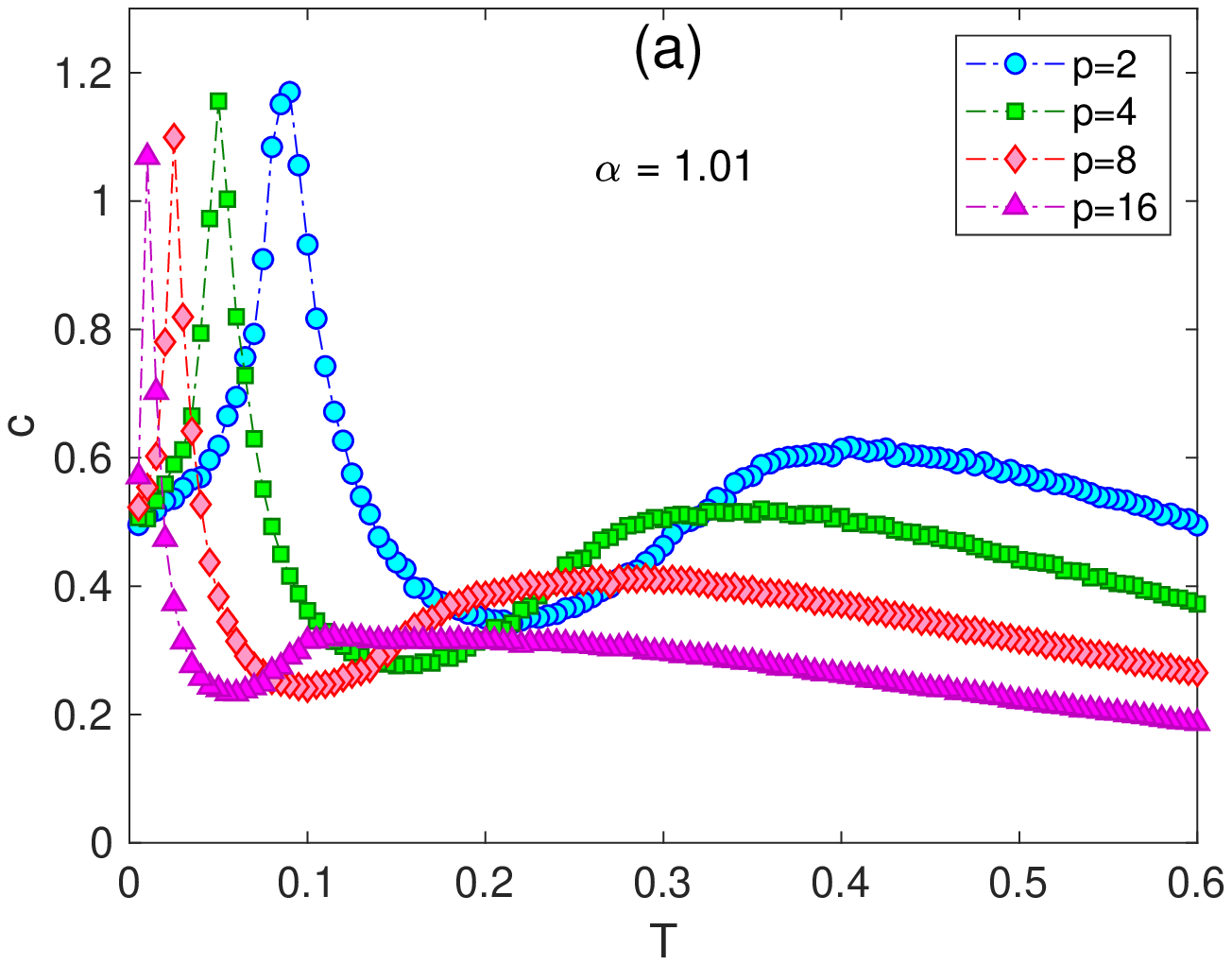}\label{fig:c-T_alp_1_01_p_even}}
\subfigure{\includegraphics[scale=0.55,clip]{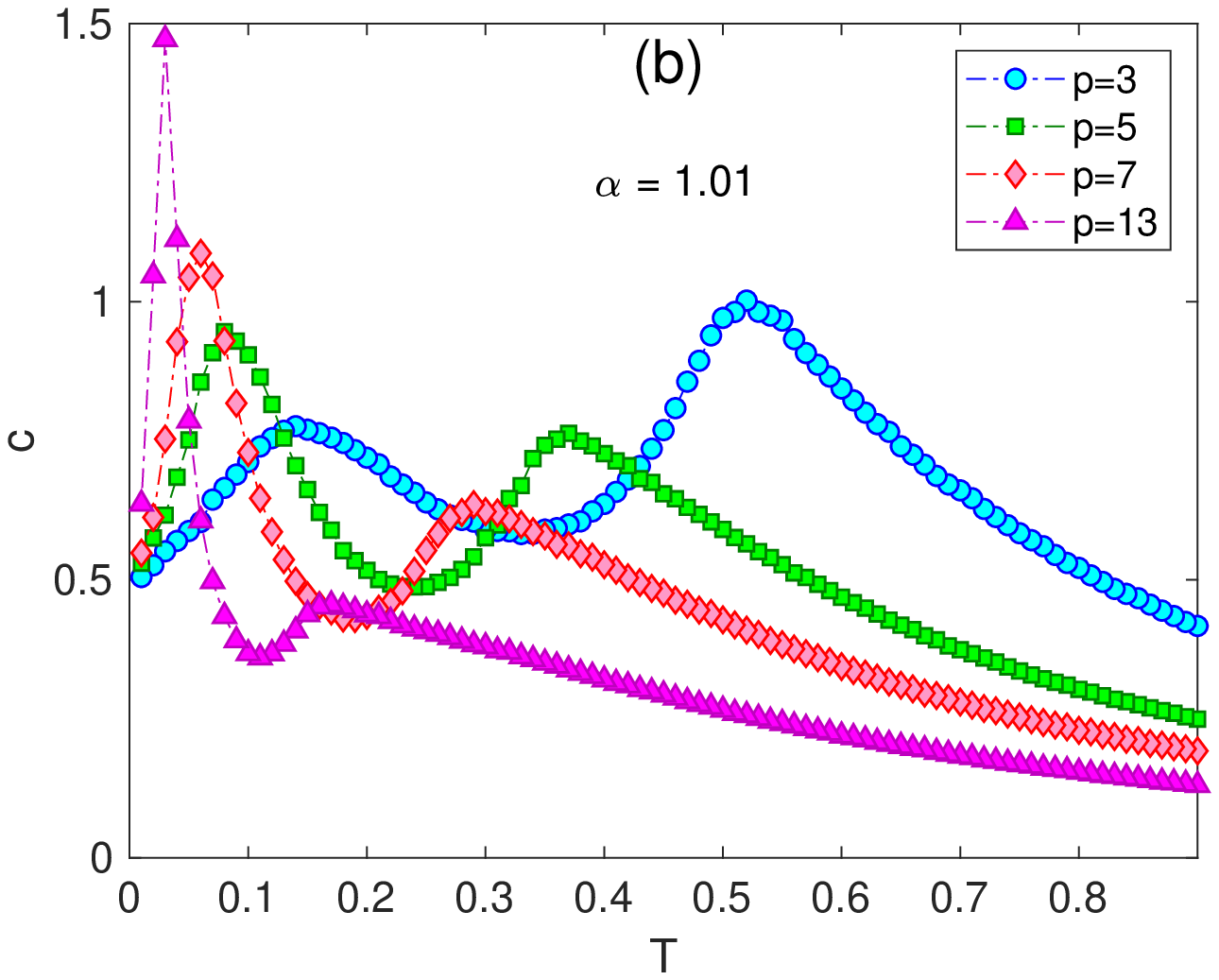}\label{fig:c-T_alp_1_01_p_odd}}
\caption{(Color online) Temperature dependencies of the specific heat for $\alpha=1.01$ and (a) even and (b) odd values of $p$.}\label{fig:c_fin}
\end{figure} 

In Fig.~\ref{fig:c_fin} we present temperature dependencies of the specific heat for a relatively small $\alpha=1.01$, in which case the effect of HOI is expected to be prominent. We separately consider the cases with even (Fig.~\ref{fig:c-T_alp_1_01_p_even}) and odd (Fig.~\ref{fig:c-T_alp_1_01_p_odd}) values of $p$. In both cases we observe two peaks, which move towards low temperatures with the increasing $p$. Nevertheless, their characteristics are somewhat different. In the former case, all low-temperature peaks are relatively sharp, while the high-temperature ones are rather broad. With the increasing $p$ the heights of all of them decrease. For odd $p$ the high-temperature peaks behave in a similar fashion, albeit their widths are narrower than for even $p$. However, the low-temperature peaks for odd $p$ are relatively broad for smaller $p$ and only become distinctly sharp for sufficiently large $p$. Moreover, unlike for even $p$, their heights increase with the increasing $p$. These features indicate the presence of low-temperature phase transitions for both even and odd $p$ but also suggest that their nature is different.

\begin{figure}[t!]
\centering
\subfigure{\includegraphics[scale=0.55,clip]{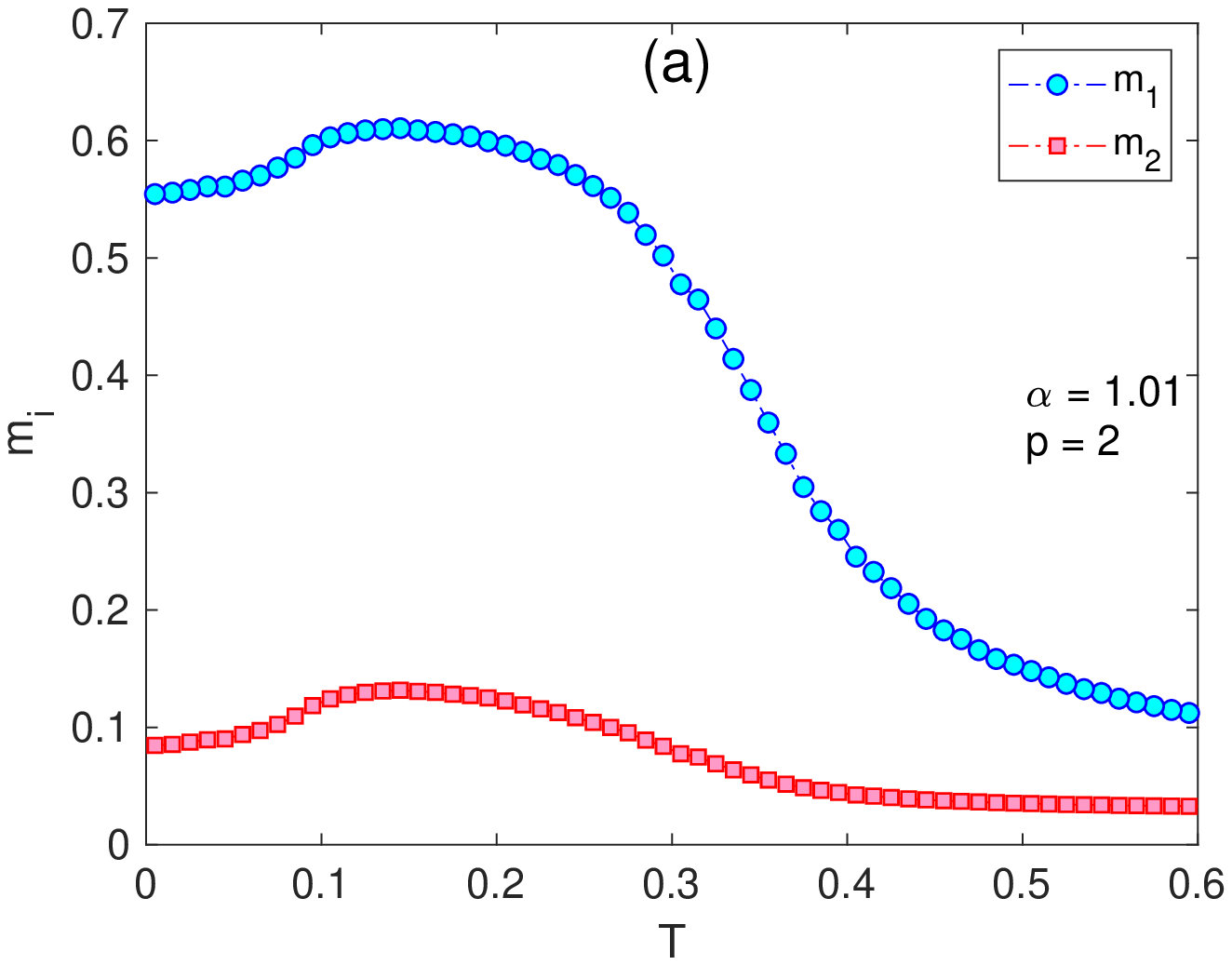}\label{fig:mi-T_alp_1_01_p_2}}
\subfigure{\includegraphics[scale=0.55,clip]{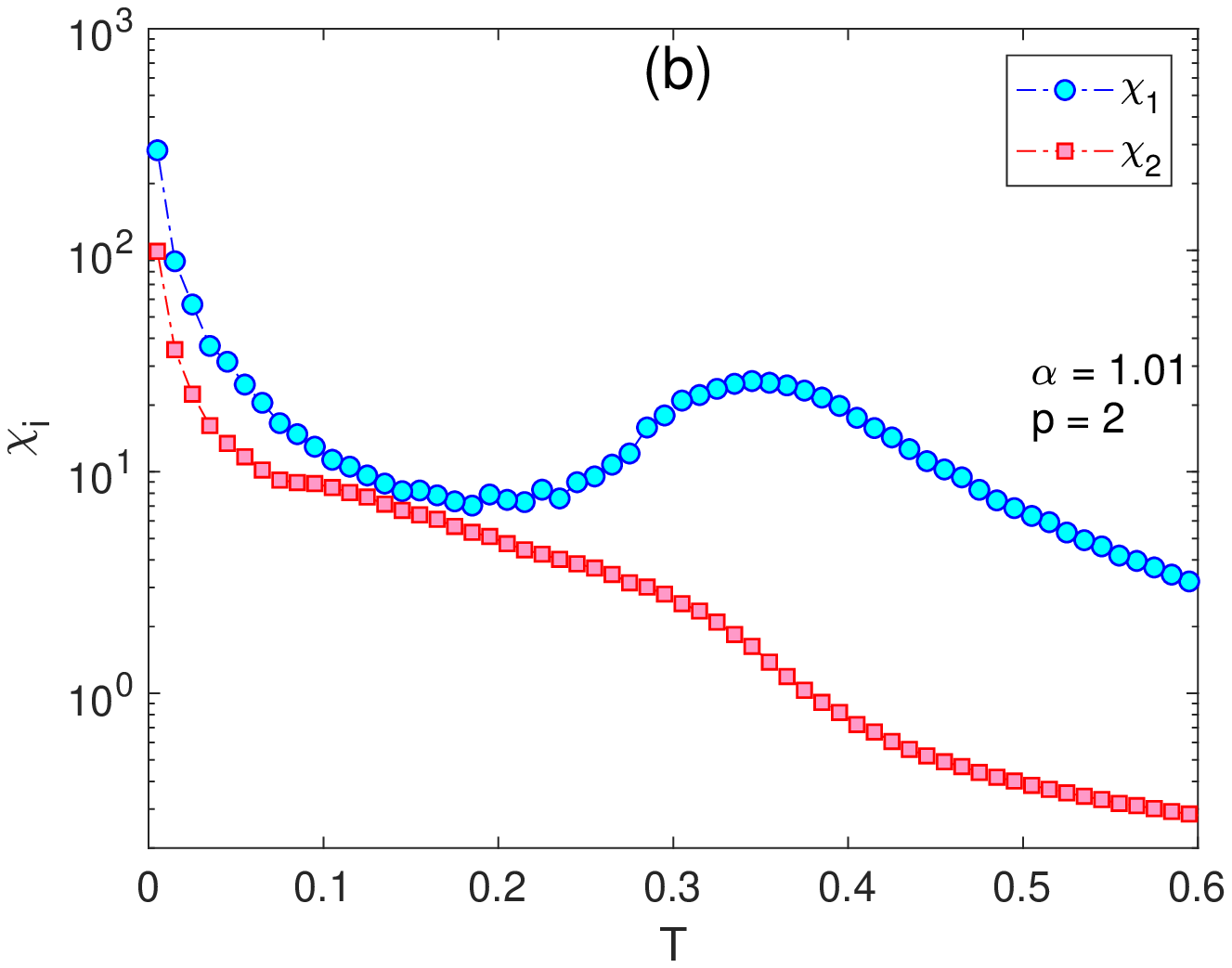}\label{fig:xii-T_alp_1_01_p_2}}\\
\subfigure{\includegraphics[scale=0.55,clip]{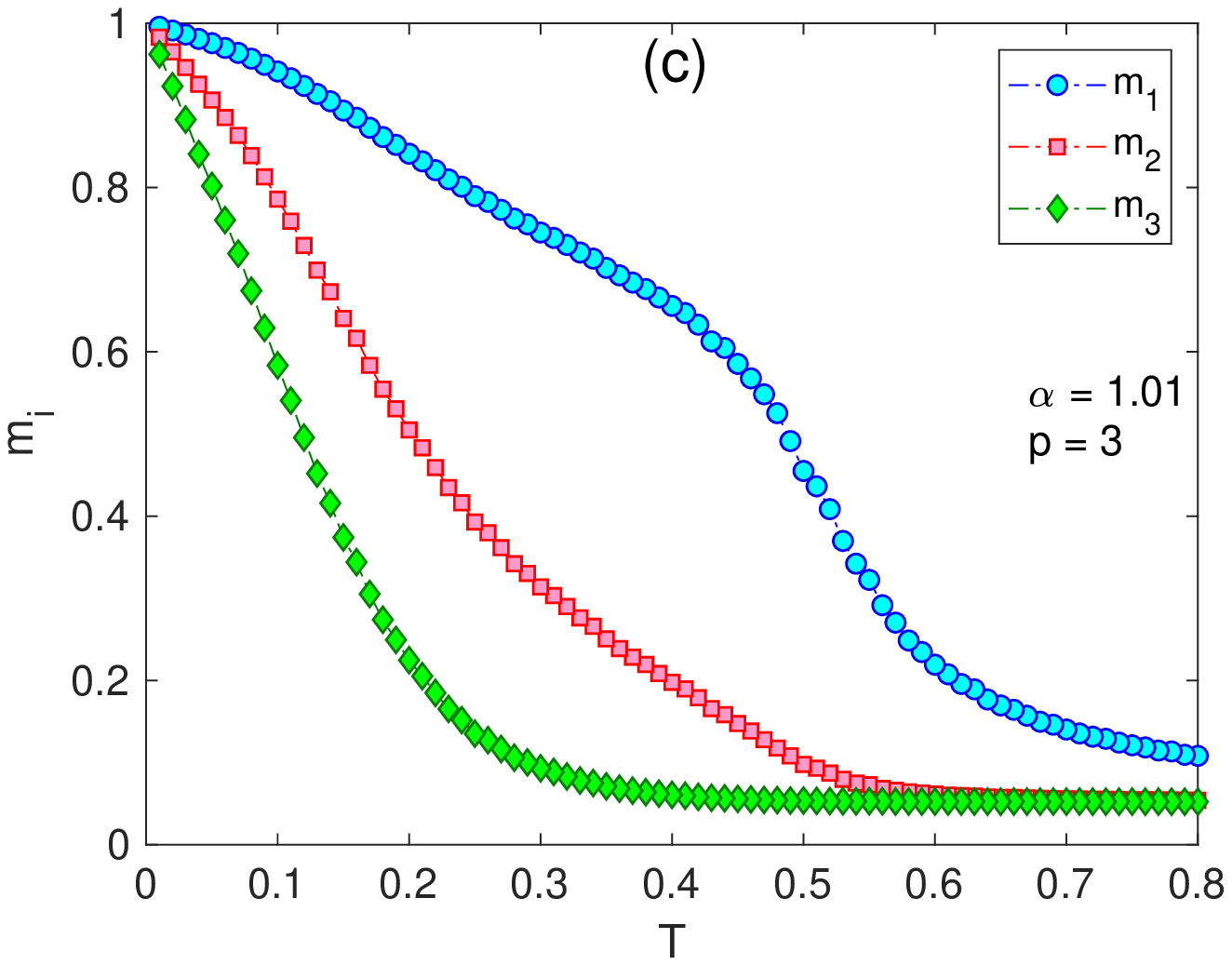}\label{fig:mi-T_alp_1_01_p_3}}
\subfigure{\includegraphics[scale=0.55,clip]{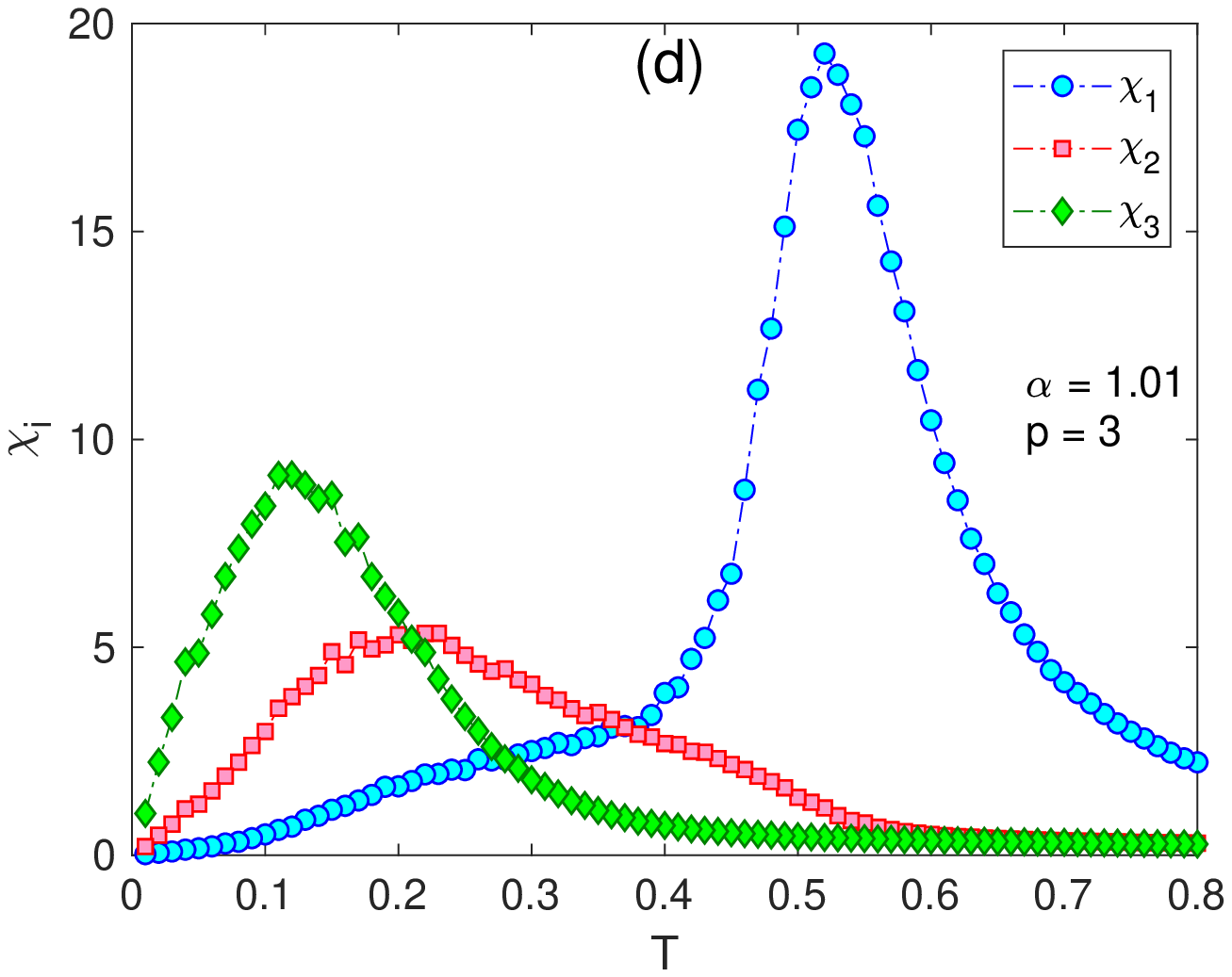}\label{fig:xii-T_alp_1_01_p_3}}
\caption{(Color online) Temperature dependencies of (a,c) the generalized magnetizations $m_i$ and (b,d) the generalized susceptibilities $\chi_i$, $i=1,\hdots,p$, for (a,b) $p=2$, (c,d) $p=3$ and the fixed value of $\alpha=1.01$.}\label{fig:mag_susc_fin}
\end{figure}

The character of the respective phases and their onsets can be revealed by studying the generalized magnetizations and the corresponding response functions, i.e., the generalized susceptiblities. In Fig.~\ref{fig:mag_susc_fin} we present them for the representative cases of even ($p=2$) and odd ($p=3$) number of the interaction terms, with $\alpha=1.01$. As the temperature is lowered, for $p=2$ the system crosses from the paramagnetic phase to a QLRO ferromagnetic phase with finite values of both $m_1$ and $m_2$. However, upon further temperature decrease, instead of approaching the ground state saturation values $m_{1,2}^{{\rm gs}}=1$, roughly at the temperature corresponding to the low-temperature specific heat peak they slightly decrease and eventually approach some nontrivial values $m_{1,2}^{{\rm gs}}<1$. The corresponding susceptibilities $\chi_{1,2}$ diverge as $T \to 0$. We note that the present $p=2$, $\alpha=1.01$ case corresponds to the previously studied XY model with the ferromagnetic-antinematic $J_1-J_2$ interactions~\cite{zuko19}, for $J_1=J_2+1=0.68$. Indeed, the results shown in Figs.~\ref{fig:c-T_alp_1_01_p_even}, \ref{fig:mi-T_alp_1_01_p_2} and~\ref{fig:xii-T_alp_1_01_p_2} can be compared with those in Fig. 4 from Ref.~\cite{zuko19}. The latter model was shown to display such a low-temperature behavior as a result of a transition to a peculiar canted ferromagnetic (CF) phase. The CF phase is characterized by highly degenerate states in which neighboring spins belonging to different sublattices are due to the coupling competition canted by a nonuniversal angle.

The situation changes dramatically by adding the third HOI term, as shown in Figs.~\ref{fig:mi-T_alp_1_01_p_3} and~\ref{fig:xii-T_alp_1_01_p_3} for $p=3$. In the competition between the $J_1$ and $J_2$ terms, striving to enforce respectively parallel and perpendicular spin arrangements, the $J_3$ term helps the former one to recover the ferromagnetic order in the low-temperature region. Consequently, for $T \to 0$ all the generalized magnetizations $m_i^{{\rm gs}} \to 1$, for $i=1,2$ and 3. Therefore, the low-temperature phase transition occurs between two different ferromagnetic phases: ${\rm F_0}$ phase appearing at intermediate temperatures and characterized by $m_{1,2}>0$ but $m_3=0$ is followed at lower temperatures by ${\rm F_1}$ phase with $m_i>0$ for $i=1,2$ and 3. The peak in $\chi_3$ roughly coincides with the low-temperature specific heat peak in Fig.~\ref{fig:c-T_alp_1_01_p_odd} and marks the ${\rm F_0}-{\rm F_1}$ phase transition.

\begin{figure}[t!]
\centering
\subfigure{\includegraphics[scale=0.55,clip]{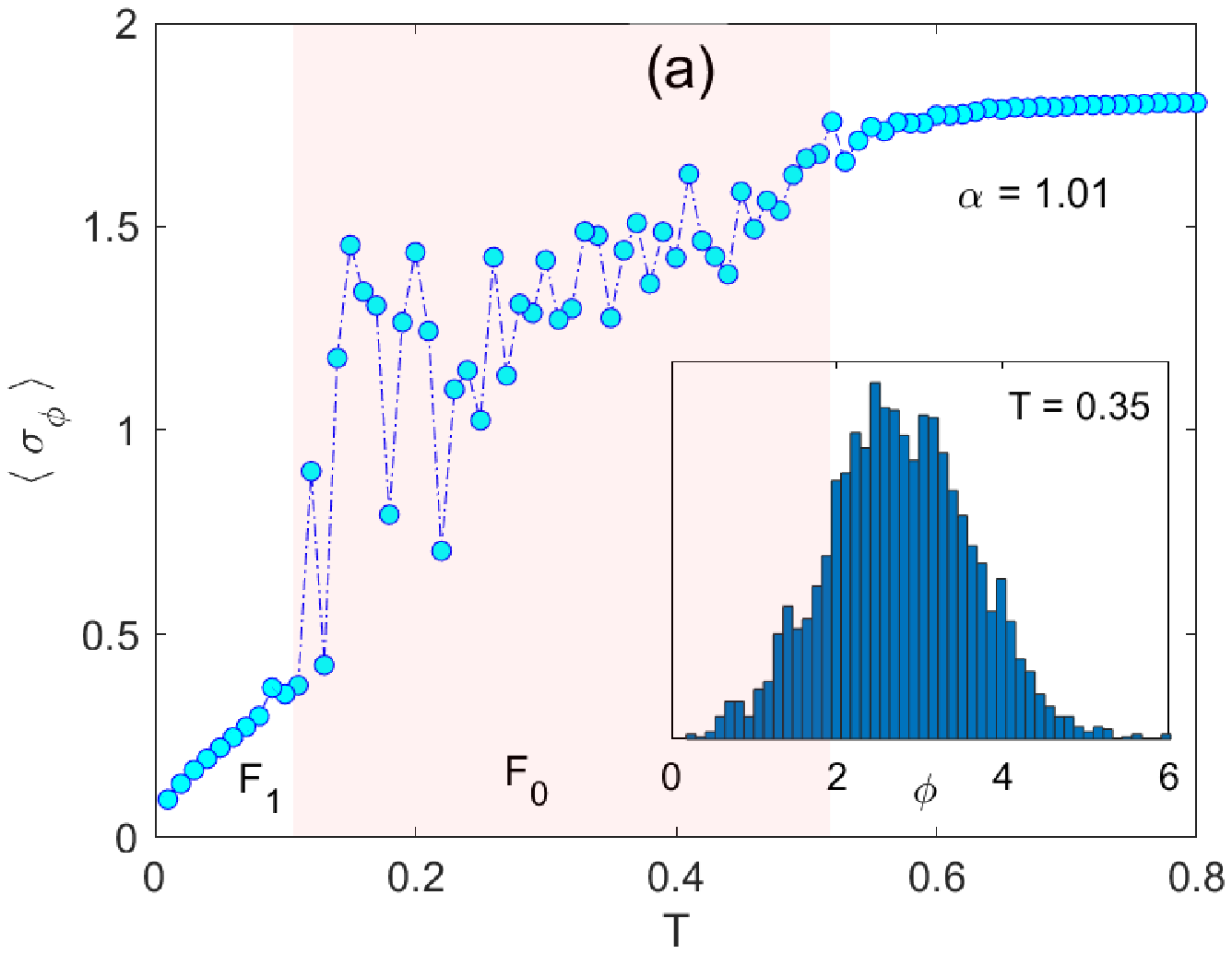}\label{fig:sigma-T_alp_1_01_p_3}}
\subfigure{\includegraphics[scale=0.55,clip]{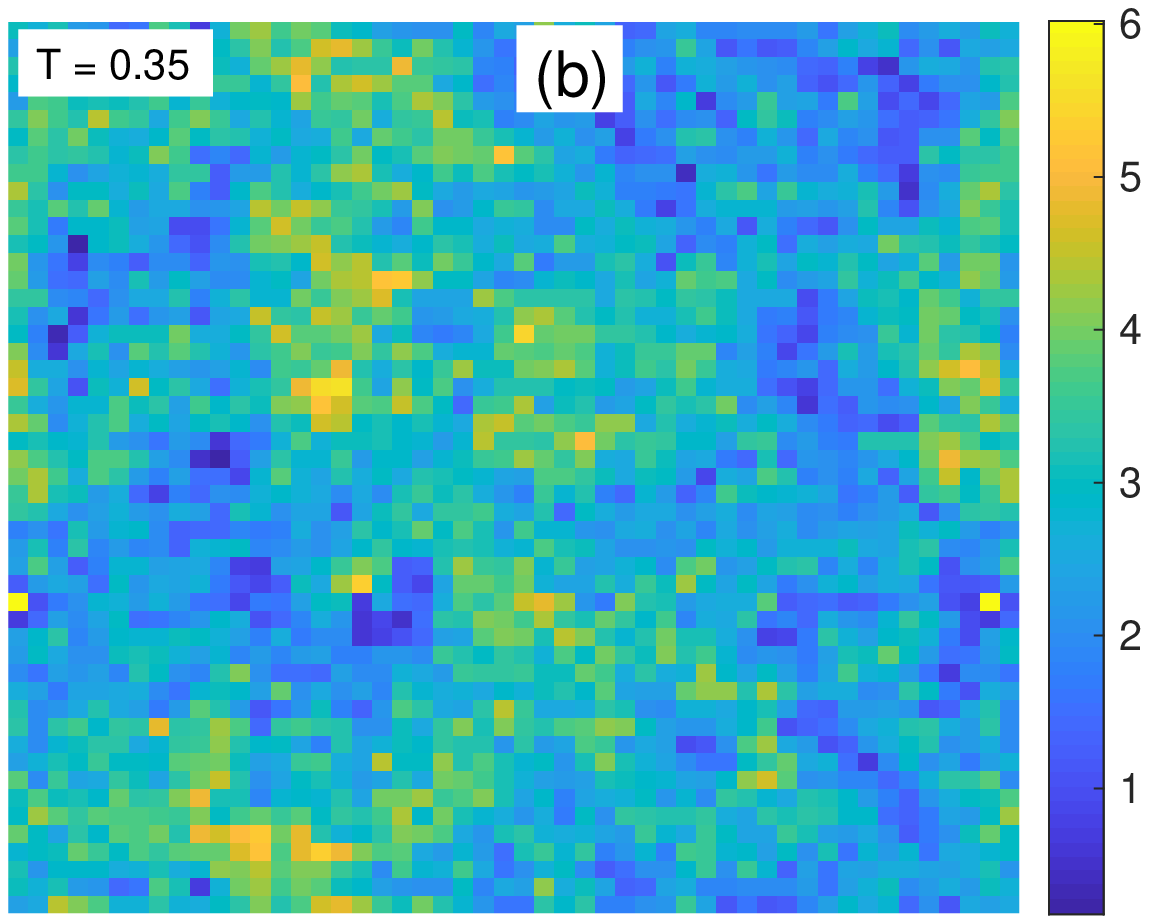}\label{fig:snap_F2_p3_alp0_01_T0_35}}\\
\subfigure{\includegraphics[scale=0.55,clip]{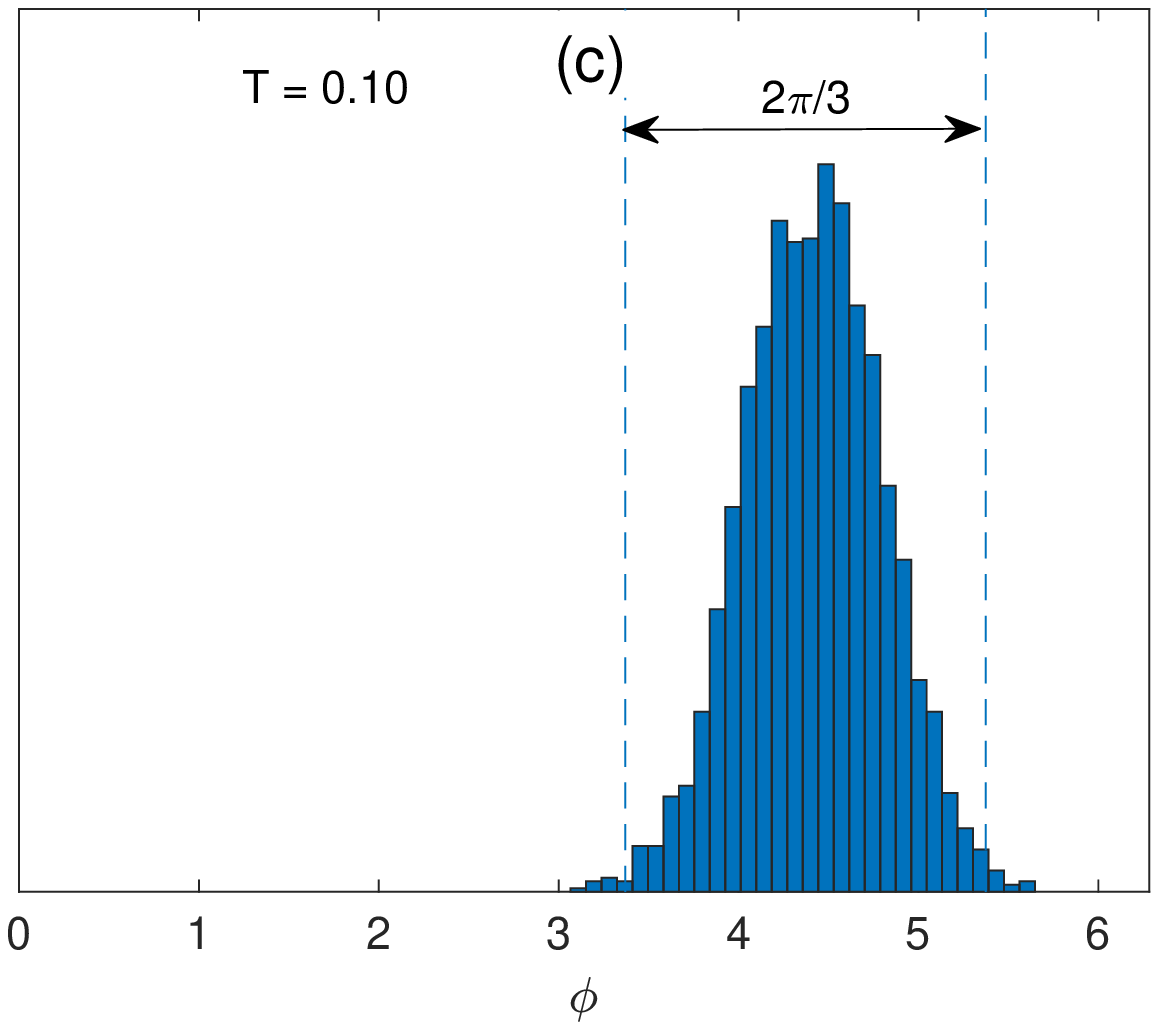}\label{fig:hist_alp_1_01_p_3_T0_1}}
\subfigure{\includegraphics[scale=0.55,clip]{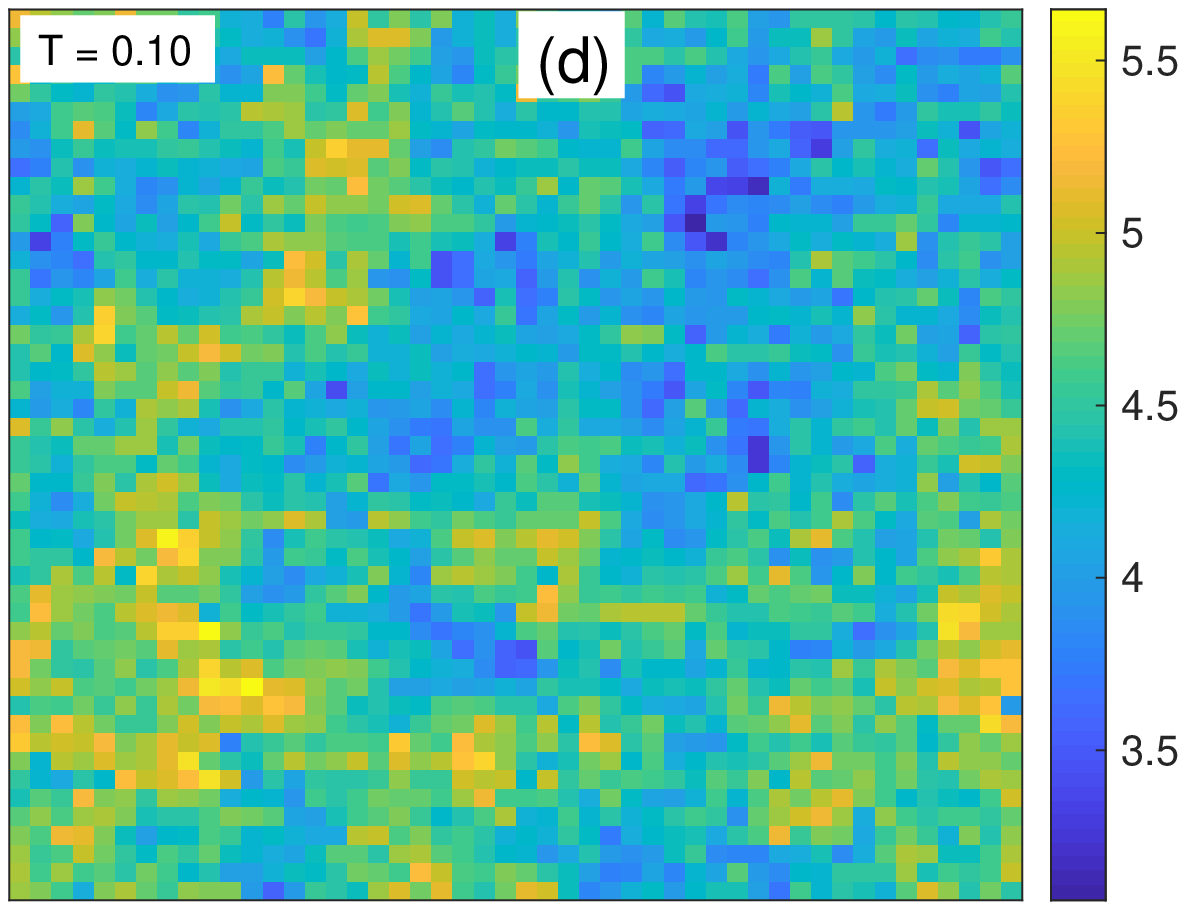}\label{fig:snap_F1_p3_alp0_01_T0_10}}
\caption{(Color online) (a) Mean value of the standard deviation of spin angle distributions as a function of temperature, for $p=3$ and $\alpha=1.01$. The inset in (a) shows the angle distribution in the phase ${\rm F_0}$ (at $T=0.35$) and the snapshot in (b) their spatial distribution on the lattice. (c) and (d) show the angle distribution and the snapshot in the phase ${\rm F_1}$ (at $T=0.10$).}\label{fig:distr_p3_alp_1_01}
\end{figure} 

Let us look into the characters of the ferromagnetic ${\rm F_0}$ and ${\rm F_1}$ phases. As mentioned above, the former one is characterized by finite $m_{1,2}$ and zero $m_3$. Fig.~\ref{fig:xii-T_alp_1_01_p_3} also shows that within ${\rm F_0}$ the susceptibilities $\chi_{1,2}$ remain anomalously elevated. As shown in Fig.~\ref{fig:distr_p3_alp_1_01}, this behavior can be related to enhanced spin fluctuations resulting from the competition between the $J_1$ and $J_2$ couplings. In Fig.~\ref{fig:sigma-T_alp_1_01_p_3} it is demonstrated by showing the mean value of standard deviations of spin angle distributions at different temperatures, for $p=3$ and $\alpha=1.01$. Evidently, within the temperature range corresponding to the ${\rm F_0}$ phase (shaded region), it shows relatively large values as well as considerable fluctuations. A representative spin angle histogram within ${\rm F_0}$ (at $T=0.35$), shown in the inset, demonstrates that the spin distribution on the lattice includes values spanning practically the entire interval $[0,2\pi]$. The corresponding spin angles snapshot is presented in Fig.~\ref{fig:snap_F2_p3_alp0_01_T0_35}. 

Below the ${\rm F_0}-{\rm F_1}$ transition temperature, i.e., within the ${\rm F_1}$ phase, the spin distribution becomes much narrower with minimal fluctuations. A typical spin angles histogram at the onset of the ${\rm F_1}$ phase (at $T=0.10$) and their spatial distribution are shown in Figs.~\ref{fig:hist_alp_1_01_p_3_T0_1} and~\ref{fig:snap_F1_p3_alp0_01_T0_10}, respectively. The histogram width in the phase ${\rm F_1}$ is reduced to about one-third of that in the phase ${\rm F_0}$. Thus, the phase ${\rm F_1}$ has similar characteristics as the phase ${\rm F_1}$ in the model ${\mathcal H}=-J_1\sum_{\langle i,j \rangle}\cos(\phi_{i,j})-J_q\sum_{\langle i,j \rangle}\cos(q\phi_{i,j})$, for $q \geq 4$~\cite{pode11,cano16} and in the present case it can be considered as a result of the interplay between the terms $J_1$ and $J_3$.

\begin{figure}[t!]
\centering
\subfigure{\includegraphics[scale=0.4,clip]{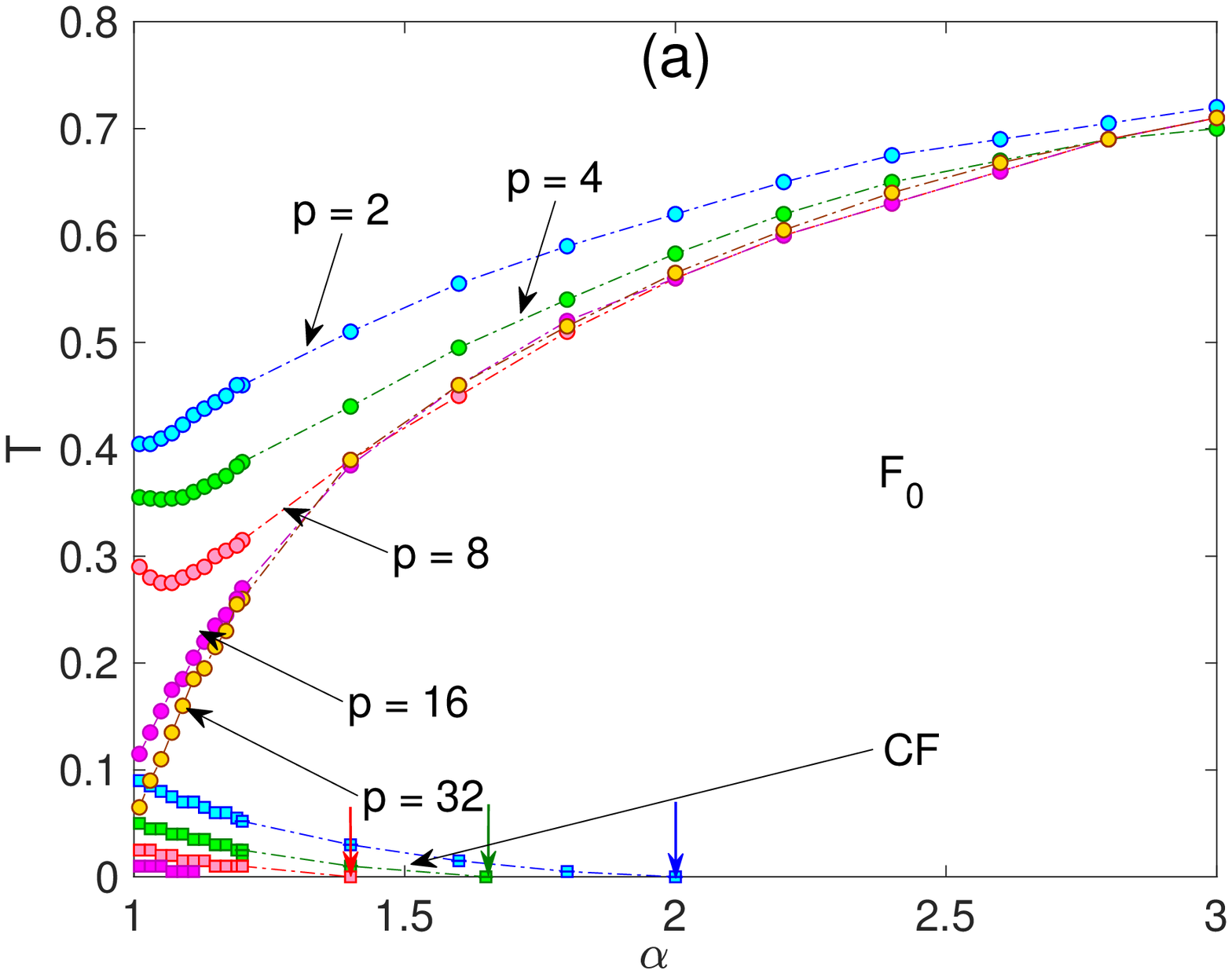}\label{fig:PD_p_even}}
\subfigure{\includegraphics[scale=0.4,clip]{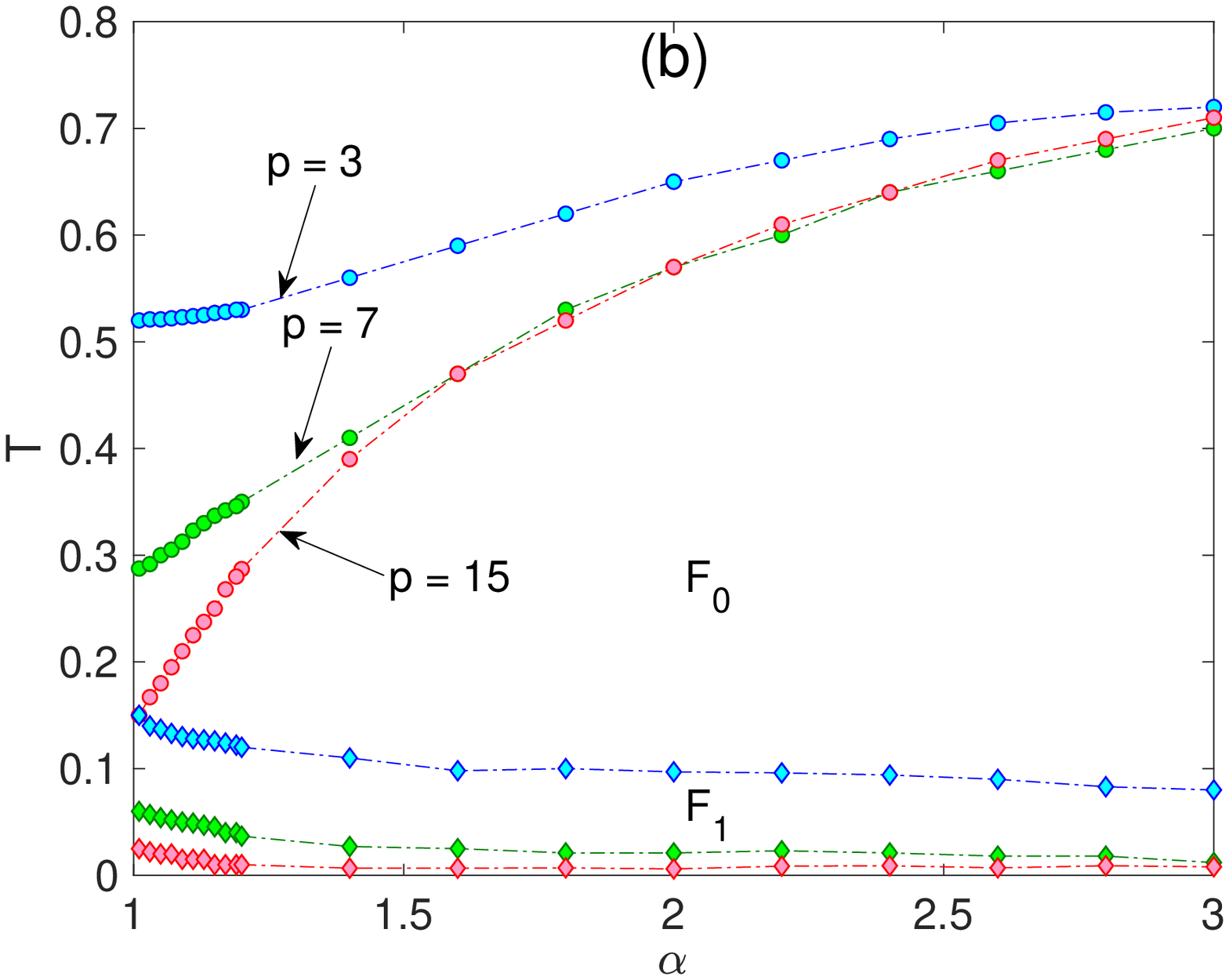}\label{fig:PD_p_odd}}
\caption{(Color online) Phase diagrams in $(\alpha-T)$ planes for (a) even and (b) odd values of $p$. ${\rm F_0}$, ${\rm F_1}$ and CF denote two different ferromagnetic and canted ferromagnetic phases, respectively. The arrows in (a) show the values of $\alpha^*$ at which CF phase terminates for different $p$.}\label{fig:PD_fin}
\end{figure}

The phase diagrams in $(\alpha-T)$ planes are presented in Fig.~\ref{fig:PD_fin}, separately for even (\ref{fig:PD_p_even}) and odd (\ref{fig:PD_p_odd}) values of $p$. For even $p$, there are either one or two phase transitions. There is always a phase transition from the paramagnetic to the ${\rm F_0}$ phase for any combination of $p$ and $\alpha$. Two phase transitions occur for the cases with $\alpha <\alpha^*$, where $\alpha^*$ is a solution of Eq.(\ref{alp}). In such cases, the ${\rm P}-{\rm F}_0$ transition is followed by the ${\rm F}_0-$CF transition, like for the $p=2$ and $\alpha=1.01$ case described above. With the increasing $p$ the transition temperatures along both ${\rm P}-{\rm F}_0$ and ${\rm F}_0-$CF phase boundaries tend to decrease, particularly for smaller $\alpha$. For a fixed value of $p$ the effect of the increasing $\alpha$ on the ${\rm P}-{\rm F}_0$ and ${\rm F}_0-$CF transition temperatures is different. Except for a small dip, observable for smaller $p$ and $\alpha$, the ${\rm P}-{\rm F}_0$ transition temperatures increase with the increasing $\alpha$. On the other hand, the ${\rm F}_0-$CF transition temperatures display a gradual decrease down to zero at $\alpha^*$. This behavior can be explained by the fact that the gradually decreasing values of $\alpha$ enforce still higher degree of competition between the odd and even HOI terms. Consequently, the increasing competition requires lower and lower temperatures to reach the ferromagnetic ${\rm F_0}$ phase and below the threshold values $\alpha^*$ also facilitates creation and gradual growth of the CF phase. 

On the other hand, for odd $p$ there are always two phase transitions for any combination of $p$ and $\alpha$ (Fig.~\ref{fig:PD_p_odd}). Another important difference is the character of the low-temperature ${\rm F_1}$ phase, as demonstrated above for $p=3$ and $\alpha=1.01$, and the fact that it seems to extend to $\alpha \to \infty$. The behavior of the transition temperatures along the ${\rm P}-{\rm F}_0$ and ${\rm F}_0-{\rm F}_1$ boundaries with the parameters $p$ and $\alpha$ is qualitatively similar to the even $p$ case, except for the absence of the small dip in the ${\rm P}-{\rm F}_0$ transition temperature curves.

\subsection{IHOI model}
The phase diagrams in Fig.~\ref{fig:PD_fin} outlined the evolution of the phase boundaries with the increasing number of HOI terms. While the high-temperature ${\rm P}-{\rm F}_0$ phase boundaries appear to consistently converge to some curve, regardless of whether $p$ is even or odd, the limiting behavior of the low-temperature ${\rm F}_0-$CF and ${\rm F}_0-{\rm F}_1$ phase boundaries is more intriguing. Below we present a MC analysis of the critical behavior of the $p \to \infty$ IHOI model, given by the Hamiltonian (\ref{Hamiltonian_inf}).

\begin{figure}[t!]
\centering
\subfigure{\includegraphics[scale=0.52,clip]{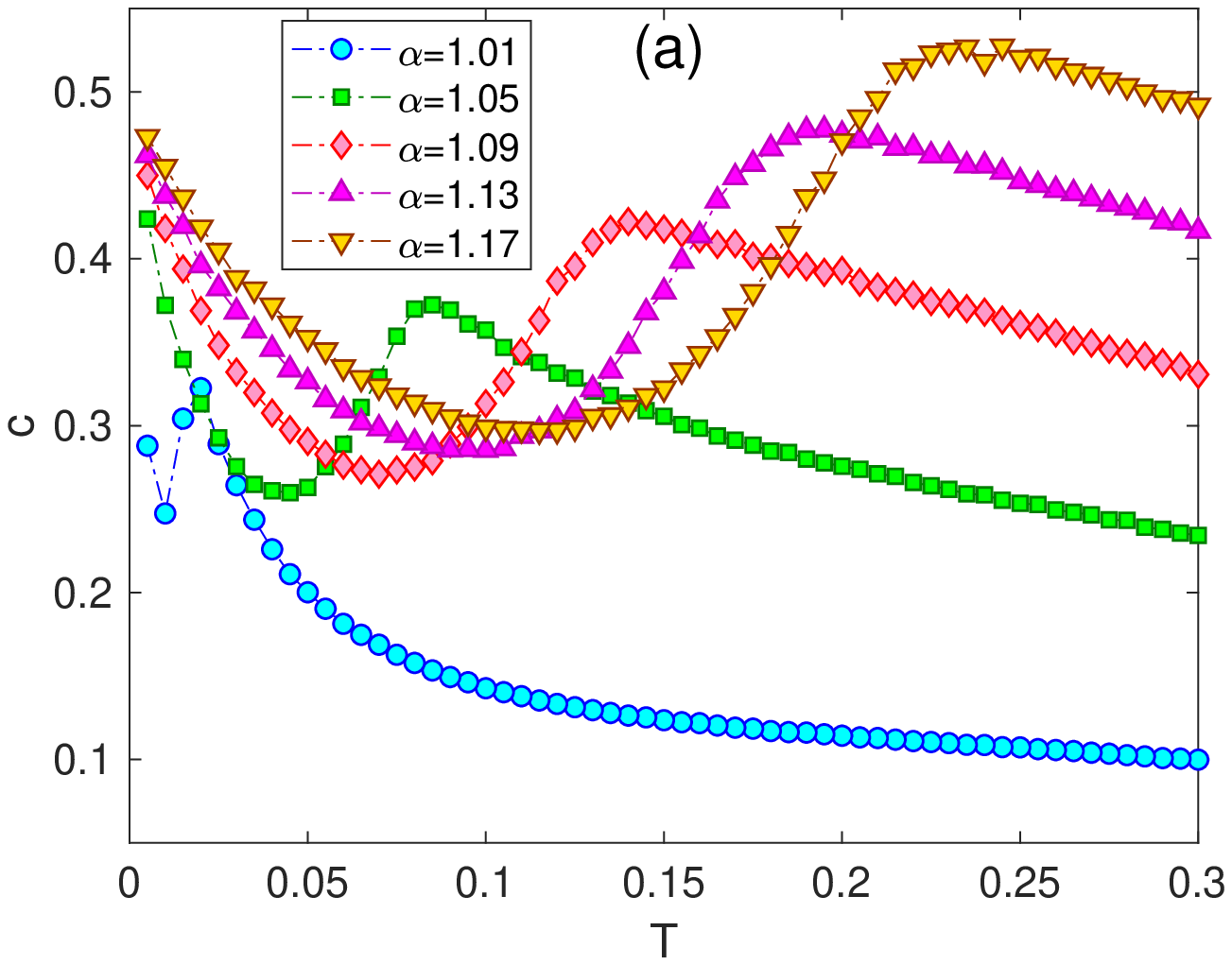}\label{fig:c-T_p_inf}}
\subfigure{\includegraphics[scale=0.52,clip]{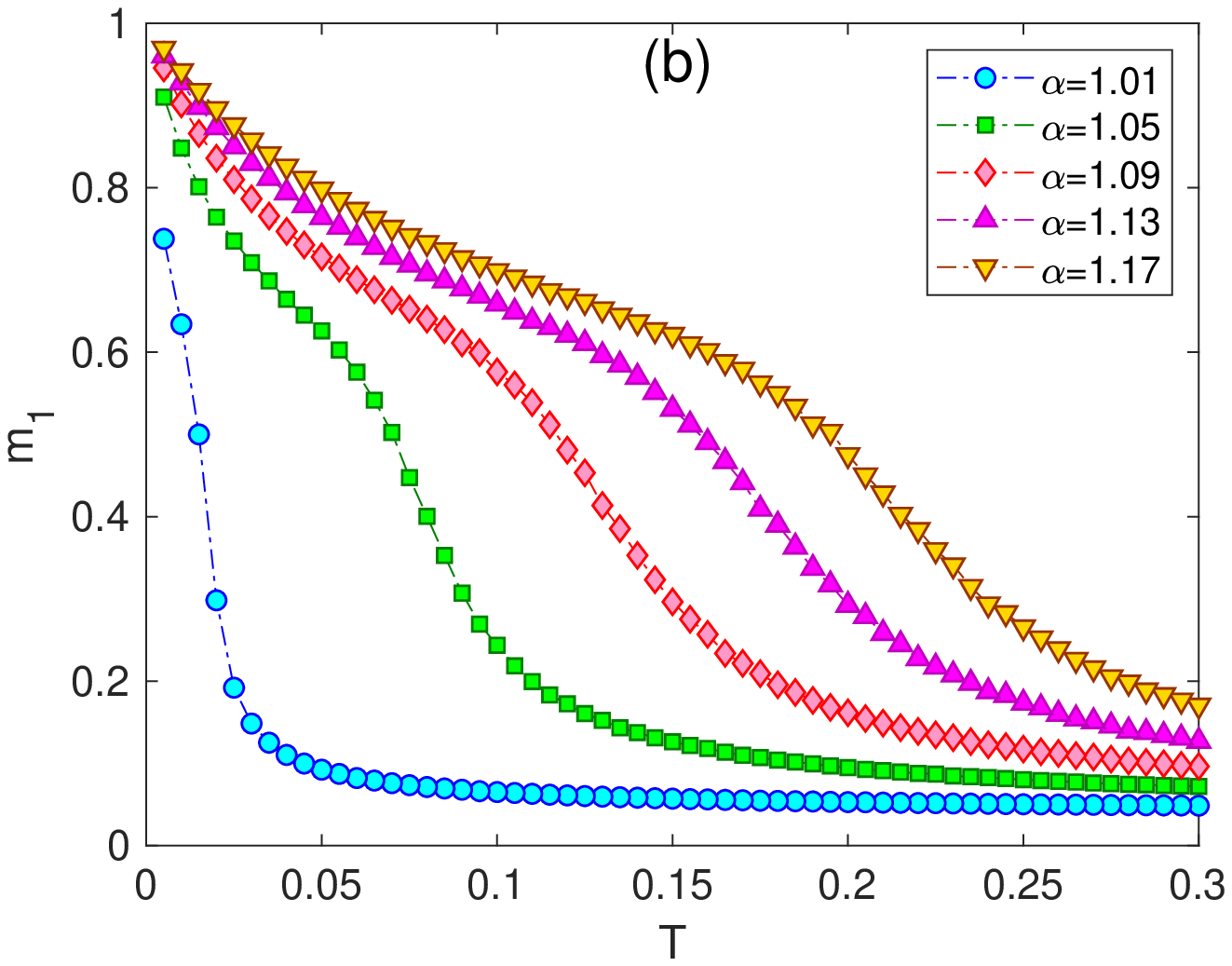}\label{fig:m1-T_p_inf}}
\caption{(Color online) Temperature dependencies of (a) the specific heat and (b) the magnetization $m_1$, for $p \to \infty$ and several values of $\alpha$.}
\label{fig:c_PD_inf}
\end{figure} 

We particularly focused on the most interesting range of relatively small values of $\alpha \to 1$. Considering the evolution of the respective phase boundaries with the increasing $p$, presented above, we are mainly interested in two aspects. Can the high-temperature ${\rm P}-{\rm F}_0$ branch be suppressed down to zero temperature for some sufficiently small values of $\alpha$, i.e., is it possible that the model would show no finite-temperature phase transition at all? Can any of the low-temperature phases (CM or ${\rm F_1}$) survive in the limit of $p \to \infty$?

In Fig.~\ref{fig:c-T_p_inf} we present temperature dependencies of the specific heat for various $\alpha$. All the curves display one peak, corresponding to the ${\rm P}-{\rm F}_0$ phase transitions, which with decreasing $\alpha$ moves towards lower temperatures. We note that the ultimate increase at $T \to 0$ is related with the convergence to $c^{\rm gs}=1/2$, expected for the XY model, and thus represents no additional anomaly. Therefore, the specific heat indicates the presence of a single ${\rm P}-{\rm F}_0$ phase transition and the absence of any other transition at lower temperatures. On the other hand, the corresponding magnetization $m_1$ curves in Fig.~\ref{fig:m1-T_p_inf} display low-temperature anomalous increase towards the saturation value of 1, which might indicate an additional phase transition. The curve for the smallest $\alpha=1.01$ suggests that there is still a phase transition occurring at finite temperature but the magnetization failed to reach the saturation value down to the lowest simulated temperature. 

\begin{figure}[t!]
\centering
\subfigure{\includegraphics[scale=0.52,clip]{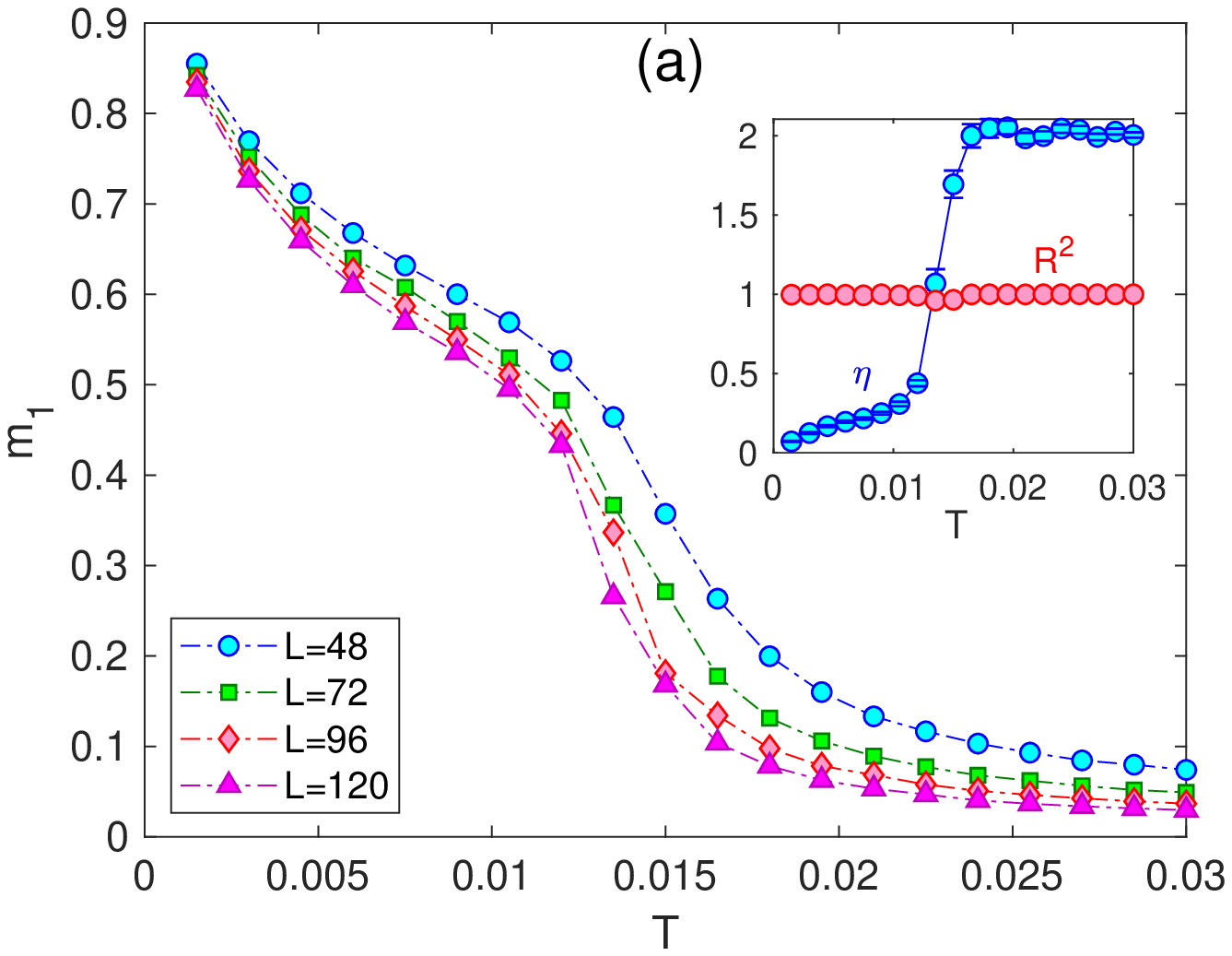}\label{fig:m1-T_p_inf_fss}}
\subfigure{\includegraphics[scale=0.52,clip]{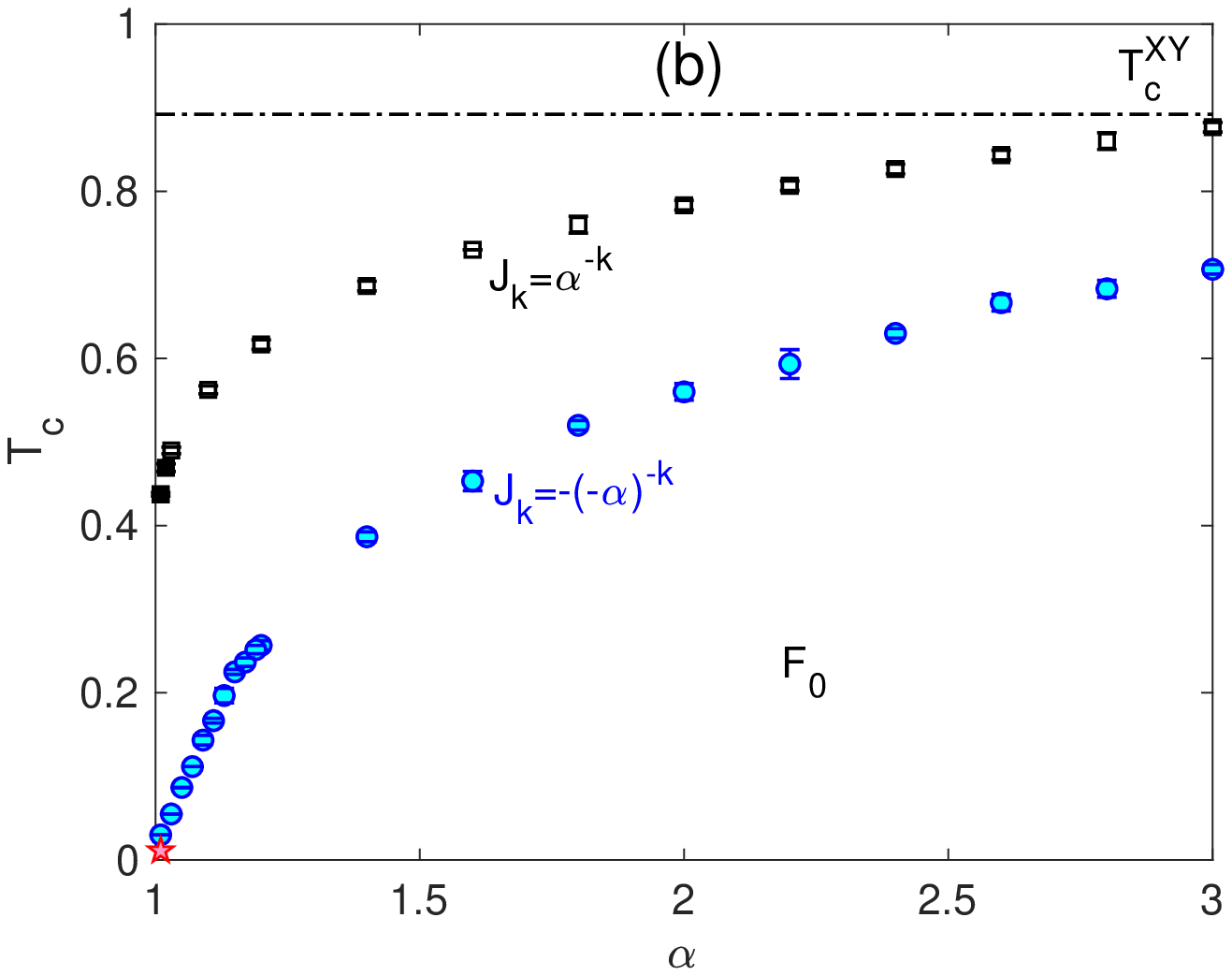}\label{fig:Tc-alp_p_inf}}
\caption{(Color online) (a) FSS behavior of $m_1$ at very low temperatures. The inset shows the temperature dependence of the correlation function critical exponent $\eta$ and the adjusted coefficient of determination $R^2$~\cite{theil61}. (b) The blue circles show the phase boundary for $p \to \infty$ as a function of the parameter $\alpha$, separating the ${\rm F_0}$ and paramagnetic phases. The (pseudo)transition temperatures are obtained from maxima of the specific heat curves shown in Fig.~\ref{fig:c-T_p_inf}. The red star represents the FSS estimation of the true value at $\alpha=1.01$. For reference, the black squares show the phase boundary for the case of only positive coupling constants $J_k$. The filled squares represent the first-order transition points and the dashed line the transition temperature of the standard XY model.}\label{fig:m1_inf}
\end{figure}

We note that the presented quantities (in both FHOI and IHOI models) correspond to a finite lattice size and thus the transition temperatures estimated from the observed anomalies are actually pseudo transition temperatures. For example, it is well known that those estimated from the specific heat peaks, have a tendency to overestimate the true values (see e.g, Ref.~\cite{zuko19}). Therefore, there is a chance that the phase boundary drops to zero at least in the limit of $\alpha \to 1$. On the other hand, based on the presented results, also the scenario of two phase transitions remains open. To verify the possibility of either of these scenarios in Fig.~\ref{fig:m1-T_p_inf_fss} we performed a FSS analysis for the lowest considered value of $\alpha=1.01$ and very low temperatures, using the scaling relation (\ref{m_FSS}). The obtained temperature dependence of the correlation function critical exponent, shown in the inset, indicates that the ${\rm P}-{\rm F}_0$ transition occurs at $T_c(L \to \infty) \approx 0.011$, which is lower than $T_c(L=48) \approx 0.03$ but still finite. The values of $m_1$ at the lowest simulated temperature neither reached the saturation value of $m_1^{\rm gs}=1$ nor they levelled off at some other value smaller than 1, as it was in the CF phase. Nevertheless, the SW theory predicts a ferromagnetic ground state with $m_1^{\rm gs}=1$ for any $\alpha>1$ but the question whether the phase transition occurs only at $T=0$ or finite temperatures remains undetermined. 

The resulting phase diagram, based on the specific heat peaks positions (blue circles) including the FSS thermodynamic limit extrapolation at $\alpha=1.01$ (red star), is presented in Fig~\ref{fig:Tc-alp_p_inf}. It shows that the ${\rm P}-{\rm F}_0$ transition temperature remains finite even in the limit of very low $\alpha \to 1$, while both low-temperature phases ${\rm F}_0-$CF and ${\rm F}_0-{\rm F}_1$, present for a finite number of HOI terms, have disappeared. We found it interesting to compare the phase boundary of the present model with the alternating HOI ($J_k=-(-\alpha)^{-k}$) signs with its uniformly ferromagnetic ($J_k=\alpha^{-k}$) counterpart~\cite{zuko17}. In the latter (black squares), inclusion of infinite number of HOI terms, besides decrease of the transition temperatures with respect to the standard XY model (dash-dotted line), led to the change to first-order transitions in the limit of $\alpha \to 1$ (filled squares). In the present case (blue circles), the reduction of the transition temperatures for a given $\alpha$ is much more pronounced and no signs of the first-order phase transitions are observed.

\section{Summary and conclusions}
We studied effects of higher-order nearest-neighbor pairwise interactions with an exponentially decreasing intensity and alternating signs, $J_k=-(-\alpha)^{-k}$, for $\alpha>1$ and $k=2,\ldots, p$, included into the standard XY model. At low temperatures, the spin wave theory predicted different behaviors for even and odd values of $p$. The low-temperature phase is characterized by a ferromagnetic QLRO with an algebraically decaying correlation function with the exponent $\eta_{\rm eff} = T/(2 \pi J_{\rm eff})$, where $J_{\rm eff}=\alpha/(1+\alpha)^2 \pm (p \alpha + p +\alpha)/(\alpha^p (1+\alpha)^2)$ for odd ($+$) and even ($-$) $p$. However, for even $p$ it is true only for sufficiently large value of $\alpha > \alpha^*$. Otherwise, the system shows a peculiar CF QLRO phase.

The phase diagrams at finite temperatures differ for even and odd values of $p$. In the former case there is a single phase transition from the paramagnetic to ferromagnetic ${\rm F_0}$ phase, if $\alpha > \alpha^*$. For $\alpha < \alpha^*$ the systems shows two phase transitions: the paramagnetic-${\rm F_0}$ transition is at lower temperatures followed by another transition to the CF phase. On the other hand, for odd values of $p$ there are always two phase transitions. In this case, the low-temperature phase is another ferromagnetic phase ${\rm F_1}$, different from ${\rm F_0}$. The character of ${\rm F_1}$ is the same as that in the model (\ref{Hamiltonian_J1-Jq}) with only $J_1$ and $J_q$ terms~\cite{cano16}. It is interesting to remark, however, that in the latter it only appears for $q > 3$. On the other hand, in the present model it is stabilized already for $q = 3$ by inclusion of the negative $J_2$ HOI. Thus, inclusion or neglecting of this intermediate term may lead to different phase diagrams.

In the limit of $p \to \infty$ the differences between even and odd values of $p$ naturally vanish, nevertheless, it is interesting to assess effects of the competition between the even and odd HOI terms of opposite signs. Compared to the model with all terms of the same positive signs in the present model the transition temperatures from the paramagnetic phase are considerably reduced. This reduction becomes the most prominent for small $\alpha$, where the two types of HOI are of comparable strengths and their mutual competition is the fiercest. In the limit of $\alpha \to 1$ the transition temperature is reduced practically to zero. On the other hand, in the model with all positive HOI the transition temperature is only reduced to about one half of that for the standard XY model and the transition character changes to the first order~\cite{zuko17}.

\begin{acknowledgments}
This work was supported by the Scientific Grant Agency of Ministry of Education of Slovak Republic (Grant No. 1/0531/19) and the scientific grants of Slovak Research
and Development Agency (Grant No. APVV-20-0150).
\end{acknowledgments}

\end{document}